\documentclass[prd,aps,superscriptaddress,nofootinbib,showpacs]{revtex4}
\usepackage{graphicx}
\usepackage{exscale}
\usepackage[intlimits]{amsmath}
\usepackage{amsfonts}
\usepackage{amssymb,amscd}
\usepackage{epsfig}
\usepackage{color}
\usepackage{pstricks}
\usepackage{rotating}
\usepackage{citesort}  

\definecolor{mygray}{gray}{.9}

\newcommand{\beqa}{\begin{eqnarray}}
\newcommand{\eeqa}{\end{eqnarray}}
\newcommand{\beq}{\begin{equation}}
\newcommand{\eeq}{\end{equation}}

\begin{document}

\title{The running coupling from the four-gluon 
vertex in Landau gauge Yang-Mills theory}

\author{Christian Kellermann}
\affiliation{Institut f\"ur Kernphysik, 
  Technische Universit\"at Darmstadt,
  Schlossgartenstra{\ss}e 9,\\ 
  D-64289 Darmstadt, Germany}
\author{Christian~S.~Fischer}
\affiliation{Institut f\"ur Kernphysik, 
  Technische Universit\"at Darmstadt,
  Schlossgartenstra{\ss}e 9,\\ 
  D-64289 Darmstadt, Germany}
\affiliation{Gesellschaft f\"ur Schwerionenforschung mbH, 
  Planckstr. 1  D-64291 Darmstadt, Germany}

\date{\today}
\begin{abstract}
We consider the running coupling from the four-gluon vertex in Landau 
gauge, SU($N_c$) Yang-Mills theory as given by a combination of dressing 
functions of the vertex and the gluon propagator. We determine these 
functions numerically from a coupled set of Dyson-Schwinger equations. 
We reproduce asymptotic freedom in the ultraviolet momentum region and 
find a coupling of order one at mid-momenta. In the infrared we find a 
nontrivial (i.e. nonzero) fixed point which is three orders of magnitude 
smaller than the corresponding fixed point in the coupling of the 
ghost-gluon vertex. This result explains why the Dyson-Schwinger and the 
functional renormalization group equations for the two point functions 
can agree in the infrared, although their structure is quite different. 
Our findings also support Zwanziger's notion of an infrared effective 
theory driven by the Faddeev-Popov determinant.
 
\end{abstract}

\maketitle

\section{Introduction}
In recent years the running coupling of Yang-Mills theory has been 
investigated in a number of approaches; for a review see 
\cite{Prosperi:2006hx}. These include lattice QCD 
\cite{Boucaud:1998bq,DellaMorte:2004bc,Kaczmarek:2005ui,
Cucchieri:2006xi,Ilgenfritz:2006gp,Allton:2007py}, analytic perturbation 
theory \cite{Shirkov:1997wi,Shirkov:2006gv}, the functional renormalization 
group \cite{Gies:2002af,Braun:2005uj,Fischer:2006vf}, Dyson-Schwinger 
equations \cite{von Smekal:1997is,Lerche:2002ep,Alkofer:2004it} and 
phenomenological extractions from experiment \cite{Brodsky:2002nb,
Baldicchi:2007zn}. The goal of these investigations is an extension of 
our knowledge of the coupling from the large momentum region towards 
small momenta of the order of $\Lambda_{QCD}$ and smaller. Perturbation 
theory alone, plagued by the problem of the Landau pole, is clearly 
insufficient for this task. In this respect it seems remarkable that the 
mere improvement of the perturbation series by analyticity constraints 
leads to a well defined running coupling that freezes out in the infrared; 
see \cite{Shirkov:2006gv} for a review of analytic perturbation theory. 

Infrared fixed points of the couplings of Yang-Mills theory have also 
been found in two functional approaches to QCD, the functional (or 'exact') 
renormalization group (FRG) and the framework of Dyson-Schwinger equations 
(DSEs); see \cite{Alkofer:2000wg,Pawlowski:2005xe,Fischer:2006ub} for 
reviews. In these approaches nonperturbative running couplings can be 
defined in terms of (gauge dependent) dressing functions of propagators 
and dressing functions of the primitively divergent vertices of the theory. 
The resulting expressions are renormalization group invariants but may be 
scheme dependent. In Landau gauge, the couplings from the ghost-gluon 
vertex, $\alpha^{gh-gl}$, the three-gluon vertex, $\alpha^{3g}$, and the 
four-gluon vertex, $\alpha^{4g}$, are given by \cite{Alkofer:2004it}:
\beqa
\alpha^{gh-gl}(p^2) &=& \frac{g^2}{4 \pi} \, G^2(p^2) \, Z(p^2) 
     \,, \label{alpha-gh-gl}\\
\alpha^{3g}(p^2) &=& \frac{g^2}{4 \pi} \, [\Gamma^{3g}(p^2)]^2 \, Z^3(p^2) 
    \,,  \label{alpha-3g}\\
\alpha^{4g}(p^2) &=& \frac{g^2}{4 \pi} \, [\Gamma^{4g}(p^2)] \, Z^2(p^2) 
     \,. \label{alpha-4g}
\eeqa
Here $g^2/4 \pi$ is the coupling at the renormalization point $\mu^2$,
whereas $Z(p^2)$ denotes the dressing function of the gluon propagator
$D_{\mu \nu}$ and $G(p^2)$ the dressing of the ghost propagator
$D^G$, i.e.
 \beq
 D^G(p^2) = - \frac{G(p^2)}{p^2} \, , \qquad
 D_{\mu \nu}(p^2)  = \left(\delta_{\mu \nu} -\frac{p_\mu 
 p_\nu}{p^2}\right) \frac{Z(p^2)}{p^2} \, .
 \eeq
The functions $\Gamma^{3g}$ and $\Gamma^{4g}$ describe the nonperturbative
dressing of the tree-level tensor structures of the three- and four-gluon
vertices. The multiplicities of the various dressing functions in 
(\ref{alpha-gh-gl})-(\ref{alpha-4g}) are related to the number of legs of 
the corresponding vertex\footnote{Note that the ghost-gluon vertex is 
finite in Landau-gauge, which explains the absence of a corresponding 
dressing function in eq.(\ref{alpha-gh-gl}). Furthermore, the bare 
four-gluon vertex is proportional to $g^2$ instead of $g$ which leads to 
factors of $\Gamma^{4g}Z^2$ instead of the naive expectation
$[\Gamma^{4g}]^2 Z^4$ from the number of legs in the vertex.}. 
The three definitions of the coupling given in 
eqs.~(\ref{alpha-gh-gl}-\ref{alpha-4g}) correspond to three different 
renormalisation schemes. The resulting couplings are related to each 
other by scale transformations and Slavnonv-Taylor identities as 
detailed e.g. in ref.~\cite{Celmaster:1979km}. In this work we focus 
on a calculation of $\alpha^{4g}(p^2)$ and compare the result with 
the previously determined coupling $\alpha^{gh-gl}(p^2)$ 
\cite{Lerche:2002ep,Fischer:2002hna}.

One of the basic ingredients to the running coupling $\alpha^{4g}(p^2)$ 
is the dressing function $\Gamma^{4g}$ of the nonperturbative four-gluon 
vertex. An evaluation of this dressing together with a corresponding 
evaluation of the gluon propagator therefore allows to study the running 
of the coupling with momentum. However, there are also other reasons why 
the nonperturbative four-gluon vertex is an interesting object. 
First of all, this vertex is the only primitively divergent one that 
allows for the formation of bound state (glueball-) poles, a phenomenon 
usually restricted to higher, superficially convergent vertices. Second, 
the vertex describes quantum corrections to elementary gluon-gluon 
scattering, which might be important e.g. for the description of 
gluon-gluon interactions in the high temperature quark-gluon plasma phase 
of QCD. Third, a number of studies indicate 
\cite{von Smekal:1997is,Lerche:2002ep,Alkofer:2004it,Zwanziger:2001kw,Zwanziger:2002ia} 
that the infrared structure of the correlation functions of Yang-Mills 
theory is connected to the confining properties of the theory via the so 
called Gribov-Zwanziger scenario. Here, long ranged correlations are 
induced in the gauge fixed theory by effects from the first Gribov horizon 
in gauge field configuration space \cite{Zwanziger:2002ia}. As we will see 
in the course of this work, the infrared behavior of the four-gluon 
vertex and the related running coupling provide additional support of this 
picture.  

This four-gluon vertex is a highly complex object due to its rich tensor 
structure generated by the four Lorentz and four color indices. As a 
consequence, this correlation function is very poorly understood so far. 
Lattice calculations of many-gluon Green's functions suffer from problems
with statistics and consequently no definite results have been obtained
so far. Within the functional continuum approach to Yang-Mills 
theory, early investigations of the vertex concentrated on the structure 
of its Dyson-Schwinger equations (see e.g. \cite{Baker:1976vz}), without 
aiming at actual solutions. Results on the one-loop level have been given 
e.g. in \cite{Celmaster:1979km,Brandt:1985zz,Papavassiliou:1992ia}. An 
attempt to solve the vertex-DSE non-perturbatively has been made in 
\cite{Stingl:1994nk,Driesen:1997wz,Driesen:1998xc} within a selfconsistent 
expansion scheme in terms of couplings and power laws of momenta. 

In this work we are going beyond these results by a combination of 
analytical and numerical methods that allow to extract the dressing 
functions of the vertex without any prejudice to their functional form. 
In section \ref{sec_trunc} we construct an approximation to the full 
Dyson-Schwinger equation of the vertex which reproduces the correct 
asymptotic behavior of the vertex as known from perturbation theory and 
infrared power counting methods \cite{Alkofer:2004it}. In section 
\ref{sec_analytic} We give analytical expressions for the vertex in these 
two limits and discuss numerical results for all momenta in section 
\ref{sec_num}. For the running coupling $\alpha_{4g}(p^2)$ we find an 
infrared fixed point, which we discuss in section \ref{sec_coupling}.
We explain why the smallness of this fixed point matches with results
from the functional renormalization group and the notion of ghost
dominance in the infrared. A summary and outlook concludes the paper.   

\section{The four-gluon vertex and its Dyson-Schwinger equation 
                                                       \label{sec_trunc}}

\subsection{Nonperturbative structure of the four-gluon vertex}

As already mentioned above, the four-gluon vertex is a highly complicated 
object with four Lorentz- and four color indices. This complexity forces 
a two step procedure: one first works with a restricted subset of possible 
combinations of Lorentz- and color tensors. This reduced complexity allows 
for a first study of the most important properties of the vertex and its 
Dyson-Schwinger equation. On the basis of these results one can then attack 
the full problem in a second step. While we report on the first part of 
this program in this work, the second part is left for future studies. Of 
course, the success of such a procedure greatly depends on the choice of the 
restricted subset. A suitable selection has been suggested in 
\cite{Driesen:1998xc} and shall also be used here.

The building blocks of the reduced tensor-structure are three Lorentz- and 
five color-tensors:
\begin{equation}
 L_{(1)}^{\kappa\lambda\mu\nu}=\delta^{\kappa\lambda}\delta^{\mu\nu}, \quad 
 L_{(2)}^{\kappa\lambda\mu\nu}=\delta^{\kappa\mu}\delta^{\lambda\nu}, \quad
 L_{(3)}^{\kappa\lambda\mu\nu}=\delta^{\kappa\nu}\delta^{\lambda\mu},
\label{lorentz}
\end{equation}
\begin{eqnarray}
 C^{(1)}_{abcd}=\delta_{ab}\delta_{cd}, \quad
 C^{(2)}_{abcd}=\delta_{ac}\delta_{bd}, \quad
 C^{(3)}_{abcd}=\delta_{ad}\delta_{bc}, \nonumber \\
 C^{(4)}_{abcd}=f_{abn}f_{cdn}, \quad C^{(5)}_{abcd}=f_{acn}f_{bdn}. \qquad
\label{colour}
\end{eqnarray}
This is the minimal subset of all possible tensor-structures, which has the
following properties \cite{Driesen:1998xc}:
\begin{itemize}
\item It is dynamically closed under DSE and Bethe-Salpeter iterations, 
      provided the only color-structure appearing in the three gluon vertex 
      is $f_{abc}$. 
\item It closes under crossing operations.
\item It contains the structure of the bare four-gluon vertex.    
\end{itemize}
The last property of this subset allows for the representation of the high 
momentum limit of the vertex in this basis and also allows for the 
calculation of the relevant dressing function for the running coupling, 
eq.~(\ref{alpha-4g}).

From these tensors a basis of the linear space of Lorentz/color-tensors is 
constructed as a direct product
\begin{equation}
 T_{(i,j);abcd}^{\kappa\lambda\mu\nu}
                  =C^{(i)}_{abcd}L_{(j)}^{\kappa\lambda\mu\nu},
\end{equation}
where we abbreviate the various combinations as follows:
\begin{equation}
\begin{tabular}{llll}
$B_{1}=L^{(1)}C_{(1)}$,  & $B_{2}=L^{(1)}C_{(2)}$,  & 
$B_{3}=L^{(1)}C_{(3)}$,  & $B_{4}=L^{(1)}C_{(4)}$,  \\
$B_{5}=L^{(1)}C_{(5)}$,  & $B_{6}=L^{(2)}C_{(1)}$,  & 
$B_{7}=L^{(2)}C_{(2)}$,  & $B_{8}=L^{(2)}C_{(3)}$,  \\
$B_{9}=L^{(2)}C_{(4)}$,  & $B_{10}=L^{(2)}C_{(5)}$, & 
$B_{11}=L^{(3)}C_{(1)}$, & $B_{12}=L^{(3)}C_{(2)}$, \\
$B_{13}=L^{(3)}C_{(3)}$, & $B_{14}=L^{(3)}C_{(4)}$, & 
$B_{15}=L^{(3)}C_{(5)}$. &
\end{tabular}
\label{blocks}
\end{equation}
Here the Lorentz- and color-indices are left implicit.
The four-gluon vertex is then represented by:
\begin{equation}
 ^{4g}\Gamma^{\kappa\lambda\mu\nu}_{abcd}\left(p_1,p_2,p_3\right)
       =\sum_{i=1}^{5}\sum_{j=1}^{3}
	\,\Gamma_{ij}\left(p_1,p_2,p_3 \right)
	T_{(i,j);abcd}^{\kappa\lambda\mu\nu},
\label{basis1}
\end{equation}
where the $T_{(i,j);abcd}^{\kappa\lambda\mu\nu}$ are elements of an 
orthonormal basis constructed from the elements $B_{1..15}$ such that 
the tree-level vertex is included. The tensors of this basis can be 
found in appendix \ref{app_B}. The algebraic manipulations involved in
the construction of this basis and also the one below have been performed 
with the use of FORM \cite{Vermaseren:2000nd}. 

Of course, the dressing functions $\Gamma_{ij}\left(p_1,p_2,p_3 \right)$ 
of the basis (\ref{basis1}) are not completely independent. Bose symmetry 
of the four external vertex legs dictates interrelations between 
combinations of the $\Gamma_{ij}\left(p_1,p_2,p_3 \right)$. This symmetry
is of course reproduced by the exact vertex-DSE, although it is far from 
trivial how this works in detail, since one external leg is always connected 
with a bare internal vertex while the others are connected with dressed
Green's functions. Thus any approximation to the full system is endangered 
to generate unsymmetric terms. These can (partly) be projected out by contraction 
with a reduced basis of tensor structures, which only include Bose symmetric 
objects. 
The construction of this reduced basis is described in appendix \ref{app_C}. 
Here we only give the result in terms of the building blocks, 
eqs.~(\ref{lorentz}),(\ref{colour}),(\ref{blocks}):
\begin{subequations}
 \begin{eqnarray}
  V_1&=&\frac{1}{108N_c^2\left(N_c^2-1\right)}
	\Big(-B_4+2B_5+2B_9-B_{10}-B_{14}-B_{15}\Big) \\
  V_2&=&\frac{1}{48N_c^4-120N_c^2+72}
	\Big(B_1+\frac{2}{3N_c}B_4
	     -\frac{4}{3N_c}B_5+B_7-\frac{4}{3N_c}B_9\nonumber \\
&&	 \hspace{3.5cm}+\frac{2}{3N_c}B_{10}+B_{13}+\frac{2}{3N_c}B_{14}
                                            +\frac{2}{3N_c}B_{15}\Big) \\
  V_3&=&\frac{1}{216\left(N_c^6-4N_c^4+N_c^2+4\right)}\nonumber \\
&&	\Big(\frac{N_c^2+6}{3-2N_c^2}B_1+B_2+B_3
 		 +\frac{2(N_c^2+1)}{3N_c-2N_c^3}B_4
		 +\frac{4(N_c^2-1)}{N_c(2N_c^2-3)}B_5+B6
		 +\frac{N_c^2+6}{3-2N_c^2}B_7 \nonumber \\
&&\hspace{2.5cm} +B_8+\frac{4(N_c^2+1)}{N_c(2N_c^2-3)}B_9
                 +\frac{2(N_c^2-1)}{3N_c-2N_c^3}B_{10}
		 +B_{11}+B_{12}\nonumber \\
&&\hspace{2.5cm} +\frac{N_c^2+6}{3-2N_c^2}B_{13}
                 +\frac{2(N_c^2+1)}{3N_c-2N_c^3}B_{14}
		 +\frac{2(N_c^2+1)}{3N_c-2N_c^3}B_{15}\Big).
 \end{eqnarray}
\label{4g-structure}
\end{subequations}
The element $V_1$ is identical to the tree-level vertex, whereas $V_2$ and 
$V_3$ represent the two only additional Bose symmetric structures that can 
be built from eq.(\ref{blocks}). The vertex is then represented by
\begin{equation}
 ^{4g}\Gamma^{\kappa\lambda\mu\nu}_{abcd}\left(p_1,p_2,p_3\right)
      =\sum_{i=1}^{3} \,^{4g}\widetilde \Gamma_{i}\left(p_1,p_2,p_3 \right) 
	V_{i;abcd}^{\kappa\lambda\mu\nu}.
\label{basis2}
\end{equation}
The object $\Gamma^{4g}(p^2)$ appearing in the running coupling 
(\ref{alpha-4g}) is then related to (\ref{basis2}) by \linebreak 
$\Gamma^{4g}(p^2) = ^{4g}\widetilde{\Gamma_{1}}\left(p_1,p_2,p_3 \right)$, 
where all external scales $p_1^2 \sim p_2^2 \sim p_3^2 \sim 
p_1\cdot p_2 \sim p_1\cdot p_3 \sim p_2\cdot p_3 \sim p^2$. We will 
come back to this coupling in section \ref{sec_coupling}. 

\subsection{The DSE for the four-gluon vertex}

Having constructed a suitable representation of the four-gluon vertex we 
now discuss the structure of its Dyson-Schwinger equation. In compact 
notation this equation reads \cite{Driesen:1998xc}: 
\begin{eqnarray}
 \parbox{3cm}
 {\includegraphics[width=3cm,keepaspectratio]{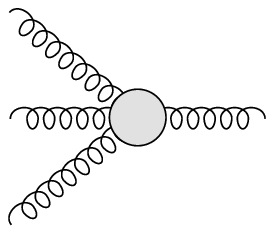}} 
 &=& \parbox{3cm}
 {\includegraphics[width=3cm,keepaspectratio]{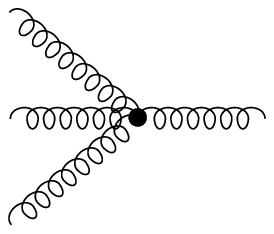}}+
 \frac{1}{2}\, \parbox{3cm}
 {\includegraphics[width=3cm,keepaspectratio]{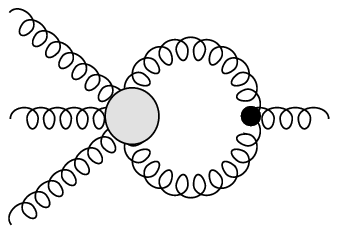}}_{(a)}
	\nonumber \\
 &-& \parbox{3cm}
 {\includegraphics[width=3cm,keepaspectratio]{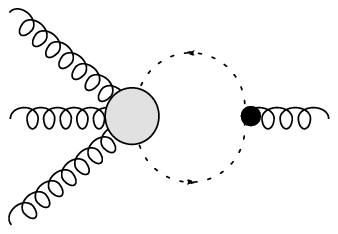}}_{(b)}
 + \frac{1}{2}\, \parbox{3cm}
 {\includegraphics[width=3cm,keepaspectratio]{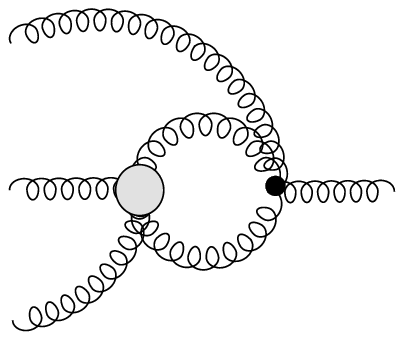}}_{(c)} 
	\nonumber \\
 &+& \frac{1}{2}\, \parbox{3cm}
 {\includegraphics[width=3cm,keepaspectratio]{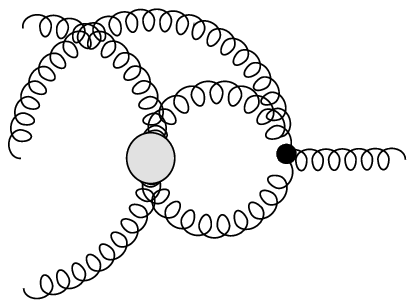}}_{(d)}+
  \frac{1}{2}\, \parbox{3cm}
  {\includegraphics[width=3cm,keepaspectratio]{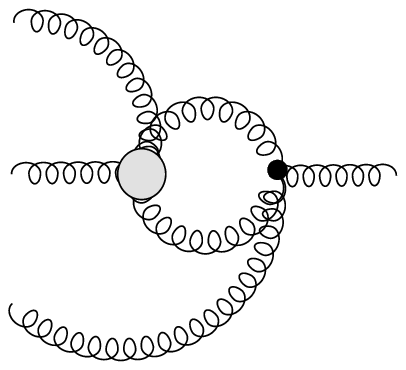}}_{(e)}
	 \nonumber \\
 &+& \frac{1}{6}\,
  \parbox{3cm}
  {\includegraphics[width=3cm,keepaspectratio]{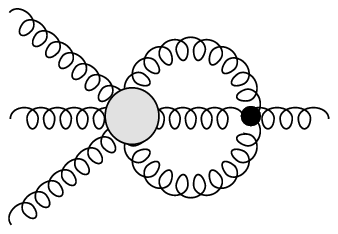}}_{(f)}\,,
\label{4g-DSE}
\end{eqnarray}
where all internal propagators are to be understood as fully dressed and
the shaded circles denote reducible vertex-functions. The decomposition of 
these functions into proper (i.e. one-particle irreducible) vertices is 
given in appendix \ref{app_D}. Here also higher n-point functions ($n=5,6$) 
appear which satisfy their own Dyson-Schwinger equations. Since in general 
one cannot solve the resulting infinite tower of equations at once, we have 
to truncate the vertex-DSE, eq.~(\ref{4g-DSE}), in a physical reasonable way. 
The truncation scheme that will be applied is as follows:
\begin{itemize}
 \item The four-gluon vertex will be reduced to the subset of structures 
       discussed above. In particular we use the Bose-symmetric 
       representation, eq.~(\ref{basis2}).
 \item The fully dressed ghost and gluon propagators in the internal
       loops are taken from their own coupled system of DSEs. These
       have been solved in \cite{Fischer:2002hna} without taking into 
       account any effects of the four-gluon vertex. By comparison with 
       lattice calculations \cite{Bonnet:2001uh,Sternbeck:2005tk} one finds 
       that this approximation in the propagator DSE leads to errors of the 
       order of ten percent in the mid-momentum region only 
       \cite{Fischer:2006ub}. The far infrared and the ultraviolet are 
       unaffected. We therefore employ these solutions in this work and
       leave an inclusion of the back-reaction of the four-gluon vertex on 
       the propagators for future studies.
 \item Due to the complexity of the four-gluon vertex DSE it seems 
       justified to reduce the number of diagrams contained in our 
       investigation to the ones that give dominant contributions in
       the infrared and ultraviolet momentum region. Since these limits
       are under analytical control (see section \ref{sec_analytic} and
       refs.~\cite{Alkofer:2004it,Fischer:2006vf}) we
       can identify these diagrams safely. In the infrared, the leading
       diagram is the ghost-loop (b), whereas in the ultraviolet leading
       contributions can be expected from all one-loop diagrams.
 \item These diagrams then contain higher n-point vertices, that will be
       reduced to two- and three-point functions using a skeleton 
       expansion (i.e. an expansion in full vertices and propagators).
 \item Selfconsistency effects of the four-gluon vertex will be neglected, 
       i.e. we drop all diagrams on the right hand side that contain
       the four-gluon vertex (e.g. the diagrams (c),(d),(e),(f) in 
       eq.~(\ref{4g-DSE}). While this approximation greatly reduces the
       complexity involved in the numerical treatment of the DSE it
       does not affect the infrared behavior of the resulting four-gluon
       vertex, since the ghost loop (b) is the dominant diagram for small 
       momenta (c.f. above). In the ultraviolet momentum region,
       however, this omission leads to a one-loop running of the vertex
       not in agreement with perturbation theory. We remedy this drawback
       by the use of an effective three-gluon vertex in diagram (b).
\item This effective three-gluon vertex  obeys the correct IR power-law 
       and generates the correct UV behavior of the four-gluon vertex 
       under absence of the diagrams (c),(d) and (e).
       A similar effective construction has been used previously in the 
       DSEs for the ghost and gluon propagators, where results close to 
       corresponding ones from lattice calculations have been obtained
       \cite{Fischer:2002hna}.
 \item The dressed ghost-gluon vertex will be replaced by the bare vertex.
       This approximation is well justified not only in the ultraviolet
       but also in the infrared momentum region. This property has already 
       been conjectured by Taylor in the early seventies \cite{Taylor:1971ff} 
       and has recently been verified numerically in continuum as well as 
       lattice calculations
       \cite{Schleifenbaum:2004id,Sternbeck:2006cg,Cucchieri:2006tf}.
       It also agrees with the all-oder analytical analysis of the DSEs 
       performed in \cite{Fischer:2006vf,Alkofer:2004it}, cf. section 
       \ref{sec:infrared}.
\end{itemize}
The resulting approximation of the Dyson-Schwinger equation of the 
four-gluon vertex then reads
\begin{equation}
 \parbox{3cm}{
	\includegraphics[width=3cm]{4GluonVertexDressed.eps}}=
 \parbox{3cm}{
	\includegraphics[width=3cm]{4GluonVertexBare.eps}}+
 perm.\frac{1}{2}\left\{\parbox{3cm}{
	\includegraphics[width=3cm]{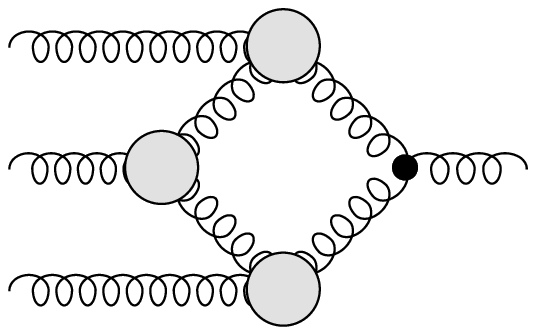}}_{|_{symm}}\right.-
 perm.\left\{\parbox{3cm}{
	\includegraphics[width=3cm]{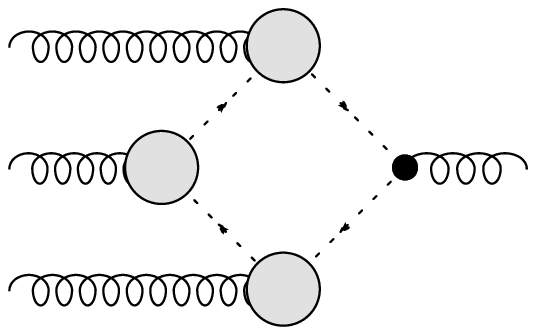}}_{|_{symm}}\right.\,.
\label{truncation}
\end{equation}\\
Here 'perm.' denotes permutations of the three external dressed legs
of the ghost-box and the gluon-box diagram. The subscript 'symm'
indicates that we average over all possible locations of the bare
vertex in the diagrams thus restoring Bose symmetry by hand. The 
dressed ghost-gluon vertices are taken bare and the dressed 
three-gluon vertex is given be the following ansatz:
\begin{equation}
 ^{3g}\Gamma^{ab}_{\lambda\mu\nu}(q,p)=
	\frac{G\left(q^2\right)^{(-\frac{1}{6}-\delta)}}
	     {Z\left(q^2\right)^{(\frac{5+3\delta}{6})}}
	\frac{G\left(p^2\right)^{(-\frac{1}{6}-\delta)}}
	     {Z\left(p^2\right)^{(\frac{5+3\delta}{6})}}
	{^{3g}_{(0)}\Gamma^{ab}_{\lambda\mu\nu}(q,p)},
	\label{3gluon}
\end{equation}
with the one-loop anomalous dimension of the ghost $\delta=-\frac{9}{44}$ 
and the dressings $G(p^2)$ and $Z(p^2)$ of the
ghost and gluon propagators. The symbol ${^{3g}_{(0)}\Gamma}$ denotes the
bare three-gluon vertex. This ansatz preserves the correct UV anomalous 
dimension of the full four-gluon vertex, as well as the correct IR power 
law of the three-gluon vertex in the scaling scenario reviewed in the next 
section.

\section{Analytical results in the infrared and ultraviolet momentum 
                                            region \label{sec_analytic}}

\subsection{Multiplicative Renormalizability}

The truncation of the DSE for the four-gluon vertex, eq.~(\ref{truncation}), 
is given explicitly by
\begin{eqnarray}
^{4g}\Gamma^{\rho\lambda\mu\nu}_{abcd}
&=& Z_4\,\, ^{4g}_{(0)}\Gamma^{\rho\lambda\mu\nu}_{abcd} \nonumber\\
&&  -  \left( \widetilde{Z}_1  g^2\int\frac{d^4q}{(2\pi)^4}
        \Gamma^{\rho} D_G \Gamma^{\lambda}
	D_G \Gamma^{\mu} D_G  \,\,{_{(0)}\Gamma^{\nu}} D_G T_{abcd} 
	\right)_{| perm.; symm} \nonumber\\
&&   + \left( Z_1 g^2\int\frac{d^4q}{(2\pi)^4} 
       {}^{3g}\Gamma^{\rho' \rho \rho''}
	D_{\rho'' \lambda'} ^{3g}\Gamma^{\lambda' \lambda \lambda''}
        D_{\lambda'' \mu'}^{3g}\Gamma^{\mu' \mu \mu''} D_{\mu'' \nu'}
        \,\,{^{3g}_{(0)}\Gamma^{\nu' \nu \nu''}} D_{\nu'' \rho'} 
	T_{abcd} \right)_{| perm.; symm}
\label{DSE1}
\end{eqnarray}
where the color factors have been subsumed in a factor 
$T_{abcd} = \left(f_{b'aa'}f_{c'bb'}f_{d'cc'}f_{a'dd'}\right)$ (recall
that we assume the ghost-gluon and three-gluon vertices to be proportional 
to $f^{abc}$ and the propagators to be diagonal in color space). The symbols 
${_{(0)}\Gamma^{\nu}}$ and ${^{3g}_{(0)}\Gamma^{\nu' \nu \nu''}}$ and 
$^{4g}_{(0)}\Gamma^{\rho\lambda\mu\nu}_{abcd}$ denote the bare 
ghost-gluon, three-gluon and four-gluon vertices respectively. All 
momentum arguments have been omitted for brevity. Note that this equation 
is already divided by a factor $g^2$ coming from the full vertex on the 
l.h.s. of eq.~(\ref{truncation}). Before we embark in the analytical analysis of this 
momentum dependence, we wish to show that this truncation scheme preserves 
multiplicative renormalizability of the four-gluon vertex DSE.

To this end we need the relations between the renormalized and the
unrenormalized but regularized Green's functions of the theory. The former 
ones are functions of the renormalization point $\mu^2$ (in addition to their 
momentum dependence), whereas the latter ones depend on the regularization scale. 
If the regularization is performed by a momentum cutoff $\Lambda$ these relations 
are given by 
\beqa
g(\mu) Z_g(\mu,\Lambda) &=& g(\Lambda) \\
D_G(p,\mu) \widetilde{Z}_3(\mu,\Lambda) &=& D_G(p,\Lambda) \\
D_{\rho\sigma}(p,\mu) {Z}_3(\mu,\Lambda) &=& D_{\rho\sigma}(p,\Lambda) \\
^{4g}\Gamma(p_i,\mu)^{\rho\lambda\mu\nu} &=& Z_4(\mu,\Lambda)
                          ^{4g}\Gamma(p_i,\Lambda)^{\rho\lambda\mu\nu}  \\
^{3g}\Gamma(p_i,\mu)^{\rho\lambda\mu} &=& Z_1(\mu,\Lambda)
                              ^{3g}\Gamma(p_i,\Lambda)^{\rho\lambda\mu} \\
      \Gamma(p_i,\mu)^{\rho} &=& \widetilde{Z}_1(\mu,\Lambda) 
                                       ^{3g}\Gamma(p_i,\Lambda)^{\rho}  
\eeqa
They are complemented by the Slavnov-Taylor identities
\beq
Z_1 = Z_g Z_3^{3/2}, \hspace{0.4cm} \tilde{Z}_1=Z_g \tilde{Z}_3 Z_3^{1/2}, 
\hspace{0.4cm} 
Z_{1F} = Z_g Z_3^{1/2} Z_2, \hspace{0.4cm}
 Z_4=Z_g^2 Z_3^2. 
\eeq
One can then analyze the dependence of the ghost and gluon box diagrams
on the renormalization point $\mu$ of the theory. We obtain for the 
renormalization point dependence of the ghost box diagram 
$_{gh}\Gamma_{abcd}^{\rho\lambda\mu\nu}(\mu^2)$
\beq 
_{gh}\Gamma_{abcd}^{\rho\lambda\mu\nu}(\mu^2) \sim 
    \frac{1}{[Z_g^2(\mu^2)]^2} [\widetilde{Z}_1(\mu^2)]^4
    \frac{1}{[\widetilde{Z}_3(\mu^2)]^4} 
 = [Z_g(\mu^2)]^2 [Z_3(\mu^2)]^3 = Z_4(\mu^2) \,,
\eeq
and for the gluon box diagram 
$_{gl}\Gamma_{abcd}^{\kappa\lambda\mu\nu}(\mu^2)$
\beq 
_{gl}\Gamma_{abcd}^{\kappa\lambda\mu\nu}(\mu^2) \sim
\frac{1}{[Z_g(\mu^2)]^2} [{Z}_1(\mu^2)]^4 \frac{1}{[{Z}_3(\mu^2)]^4} 
= [Z_g(\mu^2)]^2 [Z_3(\mu^2)]^3 = Z_4(\mu^2) \,,
\eeq
As a result, all diagrams are proportional to $Z_4(\mu^2)$, which 
guarantees the multiplicative renormalizability of the vertex DSE
in our truncation scheme.

\subsection{Yang-Mills Green's functions in the infrared \label{sec:infrared}}

The infrared behavior of the four-gluon vertex can be determined from its
Dyson-Schwinger equation by means of analytical techniques. Before we
demonstrate the details of such an analysis we have to shortly summarize
previous results on the infrared scaling of general one-particle 
irreducible Green's functions of Yang-Mills theory.

The basic idea, followed in \cite{Alkofer:2004it}, to determine the 
infrared behavior of one-particle-irreducible (1PI) Green's functions 
is to investigate their Dyson-Schwinger equations order by order in a 
skeleton expansion ({\it i.e.} a loop expansion using full propagators 
and vertices). The analysis rests upon a separation of scales, which
takes place in the deep infrared momentum region. Provided there is only 
one external momentum scale $p^2 << \Lambda_{\mathrm{QCD}}$ much smaller than 
$\Lambda_{\mathrm{QCD}}$, a self-consistent infrared asymptotic solution 
of the whole tower of Dyson-Schwinger equations for these functions is 
given by 
\beq
\Gamma^{n,m}(p^2) \sim (p^2)^{(n-m)\kappa}. \label{IRsolution}
\eeq
Here $\Gamma^{n,m}(p^2)$ denotes the dressing function of the infrared 
leading tensor structure of the 1PI-Green's function with $2n$ external 
ghost legs and $m$ external gluon legs. The exponent $\kappa$ is known 
to be positive \cite{Watson:2001yv,Lerche:2002ep}. 

A special instance of the solution (\ref{IRsolution}) are the inverse 
ghost and gluon dressing functions $\Gamma^{1,0}(p^2) = G^{-1}(p^2)$ and 
$\Gamma^{0,2}(p^2) = Z^{-1}(p^2)$, which are related to the ghost and
gluon propagators via
\beq
D^G(p^2) = - \frac{G(p^2)}{p^2} \, , \qquad
D_{\mu \nu}(p^2)  = \left(\delta_{\mu \nu} -\frac{p_\mu 
p_\nu}{p^2}\right) \frac{Z(p^2)}{p^2} \, .
\eeq
The corresponding power laws in the infrared are
\beq
G(p^2) \sim (p^2)^{-\kappa}, \hspace*{1cm} Z(p^2) \sim (p^2)^{2\kappa}\,.
\label{kappa}
\eeq
For a bare ghost-gluon vertex in the infrared, justified by lattice 
calculations \cite{Cucchieri:2006tf,Sternbeck:2006cg} and also in the 
DSE-approach \cite{Schleifenbaum:2004id}, one obtains 
$\kappa = (93 - \sqrt{1201})/98 \approx 0.595$ 
\cite{Zwanziger:2001kw,Lerche:2002ep}. Possible corrections by regular 
dressings of the vertex  in the infrared have been investigated in 
\cite{Lerche:2002ep}, where an interval $0.5 \le \kappa < 0.7$ 
has been given. Thus, although the precise value of $\kappa$ is
hitherto unknown and depends on the truncation scheme, the variation 
is quite small and not important for the results presented in this work.

An interesting consequence of the solution (\ref{IRsolution}) is the 
qualitative universality of the running coupling in the infrared.  
Renormalization group invariant
couplings can be defined from either of the primitively divergent vertices 
of Yang-Mills-theory, {\it i.e.} from the ghost-gluon vertex ($gh-gl$), 
the three-gluon vertex ($3g$) or the four-gluon vertex ($4g$) via
\beqa
\alpha^{gh-gl}(p^2) &=& \frac{g^2}{4 \pi} \, G^2(p^2) \, Z(p^2) 
     \hspace*{9mm} \stackrel{p^2 \rightarrow 0}{\sim} \hspace*{2mm} 
     {\bf const}/N_c \,, \label{gh-gl}\\
\alpha^{3g}(p^2) &=& \frac{g^2}{4 \pi} \, [\Gamma^{0,3}(p^2)]^2 \, Z^3(p^2) 
    \hspace*{2mm} \stackrel{p^2 \rightarrow 0}{\sim}
     \hspace*{2mm} {\bf const}/N_c \,,\\
\alpha^{4g}(p^2) &=& \frac{g^2}{4 \pi} \, [\Gamma^{0,4}(p^2)]^2 \, Z^4(p^2) 
    \hspace*{2mm} 
    \stackrel{p^2 \rightarrow 0}{\sim} \hspace*{2mm} {\bf const}/N_c \,.
     \label{alpha}
\eeqa
Using the DSE-solution (\ref{IRsolution}) it is easy to see that all three 
couplings approach a fixed point in the infrared. This fixed point can be
explicitly calculated for the coupling (\ref{alpha}). Employing a bare
ghost-gluon vertex one obtains $\alpha^{gh-gl}(0) \approx 8.92/N_c$
\cite{Lerche:2002ep}.

We emphasize that the eq.~(\ref{IRsolution}) solves the untruncated system 
of DSEs and the corresponding equations from the functional renormalization 
group. Thus, although $\kappa$ depends on a truncation scheme, 
(\ref{IRsolution}) does not. It is furthermore the only possible solution 
of both systems in terms of irrational power laws \cite{Fischer:2006vf}.
The resulting behaviour of the gluon and ghost propagators agree well
with the predictions deduced in the Gribov-Zwanziger and Kugo-Ojima
confinement scenarios \cite{Zwanziger:2002ia,Nakanishi:qmas}.
Nevertheless there is a caveat here: lattice Monte-Carlo simulations
have not yet been able to verify the relations (\ref{kappa}). In fact,
very recent results on large lattice indicate that the exponent of
the gluon dressing function may be close to $\kappa \approx 0.5$, 
whereas the corresponding value for the ghost dressing function may be  
considerably smaller \cite{Cucchieri:2007md,Sternbeck:2007ug}. These 
findings allow for at least two possible interpretations: they may 
indicate a different infinite volume limit than expressed by 
eq.~(\ref{IRsolution}), or they may be attributed to Gribov-copy effects 
associated with gauge fixing on large lattices. General considerations on 
the confining properties of QCD suggest the latter interpretation 
\cite{Lerche:2002ep}. Pending further clarification we will therefore
employ the behavior eq.~(\ref{IRsolution}) for the purpose of this work.  

\subsection{Infrared analysis \label{IR}}

\subsubsection{The ghost box diagram \label{IR_ghost}}

According to the general analysis of \cite{Alkofer:2004it}, the infrared
behavior of the four-gluon vertex in the presence of only one external scale
$p^2$ is given by 
\beq
\Gamma^{0,4}(p^2) \sim (p^2)^{-4\kappa}, 
\label{4gsing}
\eeq
see eq.~(\ref{IRsolution}) above. This solution is generated by the ghost
contributions to the vertex-DSE, i.e. in our truncation by the ghost box
diagram. In the following we will verify this result for one particular 
momentum configuration and determine the corresponding coefficient of the
power law. This will be useful for two reasons. First the result provides
a welcome consistency check to our numerical calculations. Second, and much
more important, together with the corresponding result for the gluon propagator 
it will give us the value for the infrared fixed point of the running coupling 
from the four-gluon vertex.
In principle, the running coupling can be calculated for every basis component
projection of the full four-gluon vertex. However, matching with perturbation
theory in the ultraviolet momentum region demands to perform this analysis
with the tree-level tensor structure, which will be done in the following.

The particular momentum configuration we choose for our analysis is given by
\begin{center}
\begin{picture}(50,80){
 \put(1,1){\includegraphics[width=3cm,keepaspectratio]{4GluonVertexDressed.eps}}}
 \put(-10,60){$p$}
 \put(-10,35){$p$}
 \put(-10,5){$p$}
 \put(100,35){$-3p$}
\end{picture}
\end{center}
It has the merit that it is invariant under permutations of the three 
dressed legs. Thus all permutations give the same results which can be
taken into account by a factor of six in front of the integral.
\footnote{At first sight one may believe that an even simpler kinematical
choice is possible, namely $p_1=p_2=-p_3=-p_4$. However, we found that such
configurations lead to results which are not stable with variation of an
numerical infrared cutoff $\epsilon$. This indicates, that the emergence of 
such a kinematic situation as a limit of a more general setup is not free of
singularities. Such 'soft' or 'collinear' singularities arise in addition
to the 'overall' singularity (\ref{4gsing}) of the four-gluon vertex. In this
work we will not touch upon these soft singularities and leave this issue 
for future studies.} 

With bare ghost-gluon vertices and projected onto the tree-level tensor
the ghost box is then given by 
\begin{eqnarray}
 \Gamma_{gh}\left(p^2\right) 
&=& -\frac{g^2N_c}{36(2\pi)^4}\int d^4q\,p^2q^2\sin^2(\theta)
	\frac{G(q+p)}{(q+p)^2}\frac{G(q+2p)}{(q+2p)^2}
	\frac{G(q+3p)}{(q+3p)^2}\frac{G(q)}{q^2} \nonumber \\
&=& -\frac{g^2N_c}{36(2\pi)^3}\int_0^{\infty}\int_0^{\pi}dq^2d\theta\,
        p^2q^4\sin^4(\theta)
	\frac{G(q+p)}{(q+p)^2}\frac{G(q+2p)}{(q+2p)^2}
	\frac{G(q+3p)}{(q+3p)^2}\frac{G(q)}{q^2}.	\nonumber \\
&& \hspace{1cm}
\label{ghostbox}
\end{eqnarray}
Note that this contribution is already Bose- symmetric in our truncation 
scheme with bare ghost-gluon vertices. The factor six from permutations
of the external legs is already included here. 

Since the internal ghost dressing functions are infrared divergent, i.e.
\begin{equation}
 G\left(p\right) = B\left(p^2\right)^{-\kappa}, \label{ghost}
\end{equation}
for $p^2 << \Lambda_{QCD}^2$
the integral is dominated by loop momenta where the internal momentum is
of the same order as the external scale $p^2$. We can thus replace the
internal ghost dressing functions by the infrared asymptotic expression
(\ref{ghost}). This leads to
\begin{equation}
 ^{IR}\Gamma_{gh}\left(p^2\right) =  
 -\frac{g^2N_cB^4}{36(2\pi)^3}\int_0^{\infty}\int_0^{\pi}dq^2d\theta\,
         p^2q^4\sin^4(\theta)
	(q+p)^{-2(1+\kappa)}(q+2p)^{-2(1+\kappa)}(q+3p)^{-2(1+\kappa)}
	q^{-2(1+\kappa)}.
\end{equation}
We then divide the momentum integration range into three parts from 
$[0,p^2]$, $[p^2,10p^2]$ and $[10p^2,\infty]$ and denote the corresponding
contributions by $I_a, I_b$ and $I_c$. 

Of course, replacing the internal ghost by the infrared asymptotic expression
(\ref{ghost}) would be a poor approximation if the contribution $I_c$ were to
dominate the total integral $I=I_a+I_b+I_c$. However, this is not the case.
$I_c$ can be evaluated using a Taylor expansion and we find its contribution
to be extremely small compared to $I_a+I_b$ provided the lower 
bound of this integral is chosen large enough. This is indeed the case for our
choice $q^2 > 10 p^2$ and we may therefore neglect $I_c$.

To evaluate the first integral, $I_a$, the approximation
\begin{equation}
 (q+p)^2(q+2p)^2(q+3p)^2\approx 36p^6\left(1+\left(\frac{q}{ap}\right)^2
 +2\frac{q}{ap}\cos(\theta)\right)^3  \label{approx_ghost}
\end{equation}
is employed, with a parameter $a>1$. This parameter can be determined 
numerically; we find $a\approx1.886$ and obtain
\begin{eqnarray}
 I_a &\approx&
	-\frac{g^2N_cB^4}{36(2\pi)^3}\int_0^{p^2}dq^2
	\frac{p^2q^4}{(36p^6q^2)^{\kappa+1}} 
	\int_0^{\pi}d\theta\frac{\sin^4(\theta)}
	{\left(1+\left(\frac{q}{ap}\right)^2
	+2\frac{q}{ap}\cos(\theta)\right)^{3(\kappa+1)}}.
\end{eqnarray}
The angular integral can be evaluated with eq.(\ref{C1}), yielding
\begin{eqnarray}
 I_a &\approx&
	-\frac{g^2N_cB^4}{36(2\pi)^3}\int_0^{p^2}dq^2\frac{p^2q^4}
	{(36p^6q^2)^{\kappa+1}} B\left(\frac{5}{2},\frac{1}{2}\right)\,
    _2F_1\left(3(\kappa+1),3\kappa+1;3;\left(\frac{q}{ap}\right)^2\right).
\end{eqnarray}
Abbreviating $z=\frac{q^2}{p^2}$ one then obtains 
\footnote{Actually eq.~(\ref{3.25}) directly shows the abovementioned soft
singularity occurring when the momentum configuration $p_1=p_2=-p_3=-p_4$
is chosen. In this case the approximation eq.~(\ref{approx_ghost}) 
becomes exact and $a=1$. However the hypergeometric function in 
eq.~(\ref{3.25}) does only converge when $|a|<1$.}
\begin{eqnarray}
 I_a &\approx& \frac{\alpha(\mu^2)\,N_c\,B^4}{192\pi}\frac{1}{(36)^{\kappa+1}}
     (p^2)^{-4\kappa}\int_0^1dz\,z^{1-\kappa}\,
     _2F_1\left(3(\kappa+1),3\kappa+1;3;\frac{1}{a^2}z\right) \nonumber\\
&=&	\frac{\alpha(\mu^2)\,N_c\,B^4}{192\pi}\frac{1}{(36)^{\kappa+1}}
\frac{\Gamma(2-\kappa)}{\Gamma(3-\kappa)}\,
	_3F_2\left(2-\kappa,3(\kappa+1),\kappa+1;3-\kappa,3;
	\frac{1}{a^2}\right) \times (p^2)^{-4\kappa}, \nonumber \\
&& \hspace{1cm}
\label{3.25}
\end{eqnarray}
with $\alpha(\mu^2)=g^2/(4\pi)$.
The last integral has been solved with the help of eq.~(\ref{C2}). 
Inserting $\kappa=(93 - \sqrt{1201})/98$ (cp. the text below 
eq.~(\ref{kappa})) and $a=1.886$ one finds
\begin{equation}
 I_a \approx 9.49\cdot 10^{-6} \cdot \alpha(\mu^2) \cdot  N_c 
 \cdot  B^4 \cdot (p^2)^{-4\kappa},
\end{equation}
which agrees with the power counting analysis, eq.~(\ref{4gsing}).

Now only the part $I_b$ where the loop momentum is of the same order of 
magnitude as the external momenta is left. It can be evaluated using a 
Chebyshev-expansion (in the loop-momentum and the polar-angle), see
appendix \ref{app-cheby} for details. Renaming variables as $x=p^2,\,y=q^2$ 
and abbreviating
\begin{eqnarray}
 f(x,y,\theta)&=&\sin^4(\theta)y^2 
        \left(x+y+2\sqrt{xy}\cos(\theta)\right)^{-(1+\kappa)}
	\left(4x+y+4\sqrt{xy}\cos(\theta)\right)^{-(1+\kappa)} \nonumber \\
&&	\times \left(9x+y+6\sqrt{xy}\cos(\theta)\right)^{-(1+\kappa)}
        y^{-(1+\kappa)} \\
g(y,\theta)&=&\sin^4(\theta)
        (1+y+2\sqrt{y}\cos(\theta))^{-(1+\kappa)}
	(4+y+4\sqrt{y}\cos(\theta))^{-(1+\kappa)} \nonumber\\
&&	\times(9+y+6\sqrt{y}\cos(\theta))^{-(1+\kappa)}y^{-(1+\kappa)},
\end{eqnarray}
and 
\beqa
 \theta_k &=& \frac{\pi}{2}\left(\cos\left(
                      \frac{(k-1/2)\pi}{N}\right)+1\right), \nonumber\\
 \tilde y_l &=& xy_k=x\left(\frac{9}{2}\cos\left(\frac{(l-1/2)\pi}{N'}\right)
                 +\frac{11}{2}\right),
\eeqa
one finds
\begin{eqnarray*}
	I_b=-\frac{g^2N_cB^4}{36(2\pi)^3}
	\frac{\pi}{N}\frac{9}{N'}(p^2)^{-4\kappa}
	\Bigg[\sum_{k=1}^N\left(
	\sum_{l=1}^{N'}g(y_l,\theta_k)+\sum_{i=2}^{N'-1}
	                 \frac{\cos(i\pi)+1}{1-i^2}
	\sum_{l=1}^{N'}\cos\left(\frac{i(l-1/2)\pi}{N'}\right)
	                  g(y_l,\theta_k)\right)\\
+	\sum_{j=2}^{N-1}\frac{\cos(j\pi)+1}{1-j^2}
          \sum_{k=1}^N\cos\left(\frac{j(k-1/2)\pi}{N}\right)\hspace{4cm}\\
\times	\left(
 	\sum_{l=1}^{N'}g(y_l,\theta_k)+\sum_{i=2}^{N'-1}
	               \frac{\cos(i\pi)+1}{1-i^2}
 	\sum_{l=1}^{N'}\cos\left(\frac{i(l-1/2)\pi}{N'}\right)
	               g(y_l,\theta_k)\right)
\Bigg].
\end{eqnarray*}
\begin{equation}
 \hspace{1cm}
\end{equation}
This expression can be evaluated numerically. It turns out that the 
expansion is well converged with $N=N'=20$. This yields
\begin{equation}
 I_b\approx 9.49\cdot 10^{-5} \cdot 
 \alpha(\mu^2) \cdot N_c \cdot B^4 \cdot (p^2)^{-4\kappa}. 
\end{equation}
This contribution is almost exactly a factor of ten larger than
$I_a$.

Putting all pieces together one finally finds
\begin{eqnarray}
 ^{IR}\Gamma_{gh}\left(p^2\right)&=& I_a+I_b+I_c \nonumber \\
&\approx& 1.04 \cdot 10^{-4}  \cdot  
\alpha(\mu^2) \cdot N_c \cdot B^4 \cdot (p^2)^{-4\kappa}.
\end{eqnarray}
This result will be used in section \ref{sec_coupling}, where we 
discuss the infrared behavior of the running coupling.

\subsubsection{The gluon box diagram}

The IR-behavior of the gluon box diagram can be estimated by power-counting 
using eq.~(\ref{IRsolution}). In this diagram, there are four gluon propagators 
along with three three-gluon-vertices. The three gluon vertices behave like
\begin{equation}
 ^{IR}\Gamma_{3g}\left(p^2\right)=C\cdot \left(p^2\right)^{-3\kappa},
\end{equation}
when $p^2\rightarrow0$. Together with the four gluon propagators, which 
contribute to the IR divergence like $C'\cdot \left(p^2\right)^{2\kappa}$ 
one gets for the gluon box diagram
\begin{equation}
 ^{IR}\Gamma_{gl}=C''\cdot\left(p^2\right)^{-\kappa}.
\end{equation}
The gluon box thus is only subleading in the infrared in agreement with our
general considerations in section \ref{sec_trunc}. Thus the coefficient 
$C''$ is of only minor interest and will not be computed here. The power law 
behaviour $(p^2)^{-\kappa}$ is well reproduced by our numerical results for
the gluon box.

\subsection{Ultraviolet analysis}
\subsubsection{The ghost box diagram}

It is known from resummed perturbation theory that in the ultraviolet 
momentum region the dressing function of the ghost propagator can be 
described by the asymptotic expression
\begin{equation}
 G(p^2)=G(\mu^2)
      \left(\omega\log\left(\frac{p^2}{\mu^2}\right)+1\right)^{\delta}, 
\end{equation}
with the one-loop anomalous dimension $\delta=-9/44$, 
$\omega=11N_c\alpha(\mu^2)/12\pi$ and some renormalization point $\mu^2$. 
This behavior is reproduced by the Dyson-Schwinger equations for the ghost 
propagator \cite{Fischer:2002hna}. Plugging this into eq.~(\ref{ghostbox}) 
and using dimensional regularization we arrive at
\begin{eqnarray}
^{UV}\Gamma_{gh}\left(p^2\right) = -\frac{g^2N_cG^4(\mu^2)}{36(2\pi)^d}
        \int d^dq\,p^2q^2\sin^2(\theta)
	\frac{\left(\omega\log\left(
	\frac{(q+p)^2}{\mu^2}\right)+1\right)^{\delta}}{(q+p)^2}
	\frac{\left(\omega\log\left(
	\frac{(q+2p)^2}{\mu^2}\right)+1\right)^{\delta}}{(q+2p)^2} 
	\nonumber \\
\quad \times 
	\frac{\left(\omega\log\left(
	\frac{(q+3p)^2}{\mu^2}\right)+1\right)^{\delta}}{(q+3p)^2}
	\frac{\left(\omega\log\left(
	\frac{q^2}{\mu^2}\right)+1\right)^{\delta}}{q^2}\,,
\end{eqnarray}
which describes the ultraviolet behavior of the tree-level projection of the 
ghost box dressing function. Since the diagram is dominated by the region 
where the loop-momentum $q^2$ is larger than the external momenta, it 
is justified to employ an angular approximation: all arguments of the 
logarithms are replaced by the loop momentum $q^2$. The integration interval 
can then be restricted to $[p^2,\infty]$. Furthermore, the denominators are 
approximated using eq.~(\ref{approx_ghost}). After evaluating the two 
trivial angular integrals and the third one using eq.(\ref{C1}) we obtain
\begin{equation}
 ^{UV}\Gamma_{gh}(p^2)=
   -\frac{g^2 N_c G^4(\mu^2)}{2592 \,(2\pi)^d}
	\frac{2\pi^{\left(\frac{d-1}{2}\right)}
	\Gamma\left(\frac{d+1}{2}\right)}
	{\Gamma^2\left(\frac{d}{2}\right)}
	\frac{1}{p^4} \int_{p^2}^{\infty}dy\, 
	y^{\frac{d}{2}-1} B\left(\frac{d+1}{2}\right)\,
	_2F_1\left(3,3-\frac{d}{2};\frac{d}{2}+1,\frac{y}{a^2x}\right)
	\left(\omega\log\left(\frac{y}{s}\right)+1\right)^{4\delta}.
\end{equation}
The hypergeometric function has the series representation
\begin{equation}
 _2F_1(\alpha,\beta,\gamma,z)=\sum_{j=0}^{\infty}
 \frac{(\alpha)_j(\beta)_j}{(\gamma)_j\,j!}z^j,
 \label{expand}
\end{equation}
with the Pochhammer symbol $(a)_j$ as introduced in appendix \ref{C3}. 
The remaining integral can be evaluated with the help of eq.~(\ref{C16}). 
We find
\begin{eqnarray}
 ^{UV}\Gamma_{gh}(p^2)&=&
	-\frac{g^2N_cG^4(s)}{1296(2\pi)^d}
	\left(\sum_{j=0}^{\infty}\frac{\pi^{\frac{d_j-1}{2}}
		\Gamma\left(\frac{d_j+1}{2}\right)}
	{\Gamma^2\left(\frac{d_j}{2}\right)}
	\frac{B\left(\frac{d_j+1}{2},\frac{1}{2}\right)}
		{a^{2j}}
	\frac{(3)_j\left(3-\frac{d_j}{2}\right)_j}
	{\left(\frac{d_j}{2}+1\right)_j\,j!}\right)
\left(\omega\log\left(\frac{p^2}{\mu^2}\right)+1\right)^{4\delta},
\end{eqnarray}\\
where the $d_j$ are different regulator dimensions, one for every 
order $j$ of the expansion eq.~(\ref{expand}). The divergence 
of the integral is absorbed into the coefficients of the logarithm
and finally cancelled by the renormalization procedure. In order
to match our results from the numerical calculations, where a momentum 
cut-off regularization will be employed, the renormalization condition 
for the analytical result is chosen such that the numerical and the 
analytical results agree at the renormalization point $\mu^2$.

As can be seen from the Slavnov-Taylor identity 
$Z_4 = Z_3/\widetilde{Z}_3^2$ the anomalous dimension of the four-gluon
vertex in the ultraviolet momentum region should equal 
$-\gamma+2\delta=1+4\delta=8/44$, where $\gamma=-13/22$ and $\delta=-9/44$ 
are the anomalous dimensions of the gluon and the ghost propagator and 
$1+2\delta+\gamma=0$. The result $4\delta=-36/44$ found here is negative 
and leads to a vanishing contribution in the ultraviolet. 
Thus the ghost box is subleading at large momenta and the leading 
contributions have to come from the gluonic diagrams.

\subsubsection{The gluon box diagram}

With the abbreviations $q_0^2=q^2$, $q_1^2=(q+p)^2$, $q_2^2=(q+2p)^2$ 
and $q_3^2=(q+3p)^2$, the gluon box integral reads
\begin{eqnarray}
 _{gl}\Gamma_{abcd}^{\kappa\lambda\mu\nu}\left(p^2\right)&=&
        \frac{g^2N_c}{72}\int\frac{d^4q}{(2\pi)^4}
	K\left(p^2,q^2,\theta\right)\frac{Z\left(q_1^2\right)}{q_1^2}
	\frac{Z\left(q_2^2\right)}{q_2^2}
	\frac{Z\left(q_3^2\right)}{q_3^2}
	\frac{Z\left(q_0^2\right)}{q_0^2} \nonumber \\
&&	\qquad\times
        \frac{\left(G\left(q_1^2\right)\right)^{(-\frac{1}{6}-\delta)}}
	{\left(Z\left(q_1^2\right)\right)^{(\frac{5+3\delta}{6})}}
	\frac{\left(G\left(q_0^2\right)\right)^{(-\frac{1}{6}-\delta)}}
	{\left(Z\left(q_0^2\right)\right)^{(\frac{5+3\delta}{6})}} 
        \left(\frac{\left(G\left(q_2^2\right)\right)^{(-\frac{1}{6}-\delta)}}
	{\left(Z\left(q_2^2\right)\right)^{(\frac{5+3\delta}{6})}}\right)^2
	\left(\frac{\left(G\left(q_3^2\right)\right)^{(-\frac{1}{6}-\delta)}}
	{\left(Z\left(q_3^2\right)\right)^{(\frac{5+3\delta}{6})}}\right)^2,
\end{eqnarray}
with the model three-gluon vertex given in eq.~(\ref{3gluon}). 
The kinematic kernel $K\left(p^2,q^2,\theta\right)$ stems from Lorentz 
contractions after projection on the tree-level vertex.  This kernel is 
complicated and lengthy and we therefore omit its explicit form. It has the 
general structure
\begin{equation}
 K\left(p^2,q^2,\theta\right)=\sum_{n=0}^{6}\cos^n\theta\times
\sum_{m=0}^{10}a_{m,n}
\left(p^2\right)^{\frac{m}{2}}\left(q^2\right)^{5-\frac{m}{2}}.
\end{equation}
For even $n$ we can replace 
$\cos^k\theta=\left(1-\sin^2\theta\right)^{\frac{k}{2}}$. If $n$ is odd, one 
has to factor out one cosine. Then the kernel has the structure
\begin{equation}
 K\left(p^2,q^2,\theta\right)
 =\sum_{n=0}^6\sum_{m=0}^{10}a_{m,n}\left(p^2\right)^{\frac{m}{2}}
	\left(q^2\right)^{5-\frac{m}{2}}\times
	\Bigg\{{\left(1-\sin^2\theta\right)^{\frac{n}{2}}
	\quad\qquad\qquad\, n\,\text{even}\atop
	\cos\theta\left(1-\sin^2\theta\right)^{\frac{n-1}{2}}
	\qquad n\,\text{odd}}
\end{equation}
This form allows using eq.~(\ref{C2}) and (\ref{C3}) for the analytic 
calculations. Similar to the ghost dressing function the ultraviolet 
behavior of the gluon dressing can be written as 
\begin{equation}
 Z(p^2)=Z(\mu^2)
 \left(\omega\log\left(\frac{p^2}{\mu^2}\right)+1\right)^{\gamma}.
\end{equation}
As before in the ghost box we employ the angular approximation 
$q_0^2=q_1^2=q_2^2=q_3^2=q^2$ in the logarithm, use the
approximation eq.~(\ref{approx_ghost}) and dimensional regularization. We then
find
\begin{equation}
 ^{UV}\Gamma_{gl}\left(p^2\right)=\frac{g^2N_c}{72(2\pi)^d}
 \frac{\left(G(\mu^2)\right)^{-1-6\delta}}
 {\left(Z(\mu^2)\right)^{1+3\delta}}
 \int d^dq\,K\left(p^2,q^2,\theta\right)
 \frac{\left(\omega\log\left(\frac{q^2}{\mu^2}\right)+1\right)^{1+4\delta}}
 {36p^6q^2\left(1+\left(\frac{q}{ap}\right)^2
 +2\frac{q}{ap}\cos(\theta)\right)^3}.
\end{equation}
Evaluating the trivial angular integrations and again restricting the
integral on the interval $[p^2,\infty]$ then leads to
\begin{eqnarray}
 ^{UV}\Gamma_{gl}(p^2)&=&\frac{g^2N_c}{72(2\pi)^d}
 \frac{2\pi^{\left(\frac{d-1}{2}\right)}\Gamma\left(\frac{d+1}{2}\right)}
 {\Gamma^2\left(\frac{d}{2}\right)}
 \frac{\left(G(\mu^2)\right)^{-1-6\delta}}
 {\left(Z(\mu^2)\right)^{1+3\delta}} \nonumber \\
 &&\times\int_{p^2}^{\infty}dy\int_{0}^{\pi}d\theta\,
 (q^2)^{\frac{d-3}{2}}\sin^{d-2}{\theta}
 \frac{K(p^2,q^2,\theta)}
 {36x^3\left(1+\frac{q^2}{a^2x}
 +2\sqrt{\frac{(q^2)}{a^2x}}\cos(\theta)\right)^3}
 \left(\omega\log\left(\frac{q^2}{\mu^2}\right)+1\right)^{1+4\delta} 
\end{eqnarray}
with eqs.(\ref{C2}),(\ref{C3}) and (\ref{C16}) one then finds
\begin{equation}
 ^{UV}\Gamma_{gl}(p^2)=C \,\cdot\,
 \left(\omega\log\left(\frac{p^2}{\mu^2}\right)+1\right)^{1+4\delta},
\label{UVglue}
\end{equation}
with a regularized factor $C$. This divergent factor is extremely lengthy 
and we therefore refrain from giving it explicitly here. However we note that
the divergence is such that it matches the one of $Z_4$ in the four-gluon DSE
thus guaranteeing a finite result for the four-gluon vertex on the left hand 
side of the DSE. The momentum dependence of (\ref{UVglue}) and in
particular the anomalous dimension $1+4\delta$ of the logarithm is in 
agreement with the expectations from the Slavnov-Taylor identity for the 
four gluon vertex renormalization factor as discussed above. 

\section{Numerical results for the four-gluon vertex \label{sec_num}}

\subsection{Numerical methods}
For the numerical investigation the subtracted version of 
eq.~(\ref{truncation}), is considered.
\begin{eqnarray}
 \parbox{2cm}{
	\includegraphics[width=2cm]{4GluonVertexDressed.eps}}\Bigg|_{p^2}&=&
 \parbox{2cm}{
	\includegraphics[width=2cm]{4GluonVertexDressed.eps}}\Bigg|_{\xi^2}+
 perm.\frac{1}{2}\cdot\left(\parbox{2cm}{
	\includegraphics[width=2cm]{GluonBox.eps}}\Bigg|_{p^2}-
	\parbox{2cm}{
	\includegraphics[width=2cm]{GluonBox.eps}}\Bigg|_{\xi^2}\right)_{symm.}
	\nonumber \\
&&\hspace{2.5cm}\,-perm\cdot\left(\parbox{2cm}{
	\includegraphics[width=2cm]{GhostBox.eps}}\Bigg|_{p^2}-
	\parbox{2cm}{
	\includegraphics[width=2cm]{GhostBox.eps}}\Bigg|_{\xi^2}\right)_{symm.},
\label{sub_truncation}
\end{eqnarray}
with some subtraction point $\xi^2$, which is set equal to the ultraviolet 
momentum cutoff $\Lambda^2=10^{10}\text{ GeV}^2$ for reasons of numerical
stability. Furthermore we also introduce an infrared cutoff $\epsilon^2$, 
which is chosen as $\epsilon^2=10^{-10}\text{ GeV}^2$. The numerical 
integration is carried out using a Gau\ss-Legendre algorithm on a 
logarithmic grid. 

In eq.~(\ref{sub_truncation}), the renormalization constant $Z_4$ stemming 
form the tree-level vertex after projection, is replaced by an integral and
the full dressed vertex function at the subtraction point. Its value has to 
be fixed by renormalization. This will be done such that the running coupling 
from the four-gluon vertex has the same value as the running coupling from 
the ghost-gluon vertex at the renormalization point $\mu^2$. As input we use
the value $\alpha(\mu^2 = 1.713\text{ GeV}^2)=0.97$ determined in 
\cite{Fischer:2002hna} within a momentum subtraction scheme.

Before we present our results we need to discuss two further technical points:
\begin{itemize}
\item We explicitly checked that our results are independent of the 
ultraviolet and infrared cutoffs. This is indeed the case, if these
cutoffs are at least three orders of magnitude larger/lower than the 
largest/lowest of the external momenta. In addition, due to the
complicated kinematics of the gluon box we had to use quite a large 
number of sampling points for the radial momentum integral (typically
5000 points on a logarithmic grid). To further improve the numerical 
accuracy the momentum integral has been split into three parts, 
integrating the infrared up to a small region $p^2\pm\Delta p^2$ around 
the external scale, the small region itself and then up to infinity. 
In the numerical calculations $\Delta p^2$ is chosen as 
$\Delta p^2=0.01p^2$. In the numerical angular integral we find that
is is numerically advantageous to integrate over the cosine of the 
angle.
\item The crucial assumption in our infrared analysis of the vertex-DSE 
was, that the integrand of the diagrams are dominated by loop momenta of 
the order of the (small) external momentum. This we verified explicitly also 
numerically.
\end{itemize}
As input for the ghost- and gluon propagators we take the following analytical
expressions
\begin{equation}
 \alpha(x)=\frac{\alpha(0)}
 {\ln[e+a_1(x/\Lambda^2_{QCD})^{a_2}+b_1(x/\Lambda^2_{QCD})^{b_2}]}; 
 \hspace*{2cm}
 R(x)=\frac{c(x/\Lambda^2_{QCD})^{\kappa}+d(x/\Lambda^2_{QCD})^{2\kappa}}
	{1+c(x/\Lambda^2_{QCD})^{\kappa}+d(x/\Lambda^2_{QCD})^{2\kappa}}
\end{equation}
\begin{equation}
 Z(x)=\left(\frac{\alpha(x)}{\alpha(\mu)}\right)^{1+2\delta}R^2(x); 
 \hspace*{3cm}
 G(x)=\left(\frac{\alpha(x)}{\alpha(\mu)}\right)^{-\delta} R^{-1}(x),
\label{input}
\end{equation}
with parameters
\begin{center}
\begin{tabular}{|c|c|c|c|c|c|c|c|c|}
 \hline
 $\alpha(0)$ & $\alpha(\mu)$ & $a_1$   & $a_2$   & $b_1$   & $b_2$   
 & $c$     & $d$     & $\Lambda_{QCD}$ \\\hline
 $8.915/N_C$ & $0.97$        & $1.106$ & $2.324$ & $0.004$ & $3.169$ 
 & $1.269$ & $2.105$ &$ 0.714 \textnormal{GeV}$ \\\hline
\end{tabular}
\end{center}
and the anomalous dimension $\delta=-9/44$ of the ghost. These expressions
have been fitted to the numerical results of \cite{Fischer:2002hna}
for the coupled system of DSEs for the ghost- and gluon propagators.

\subsection{Results}

\begin{figure}[t]
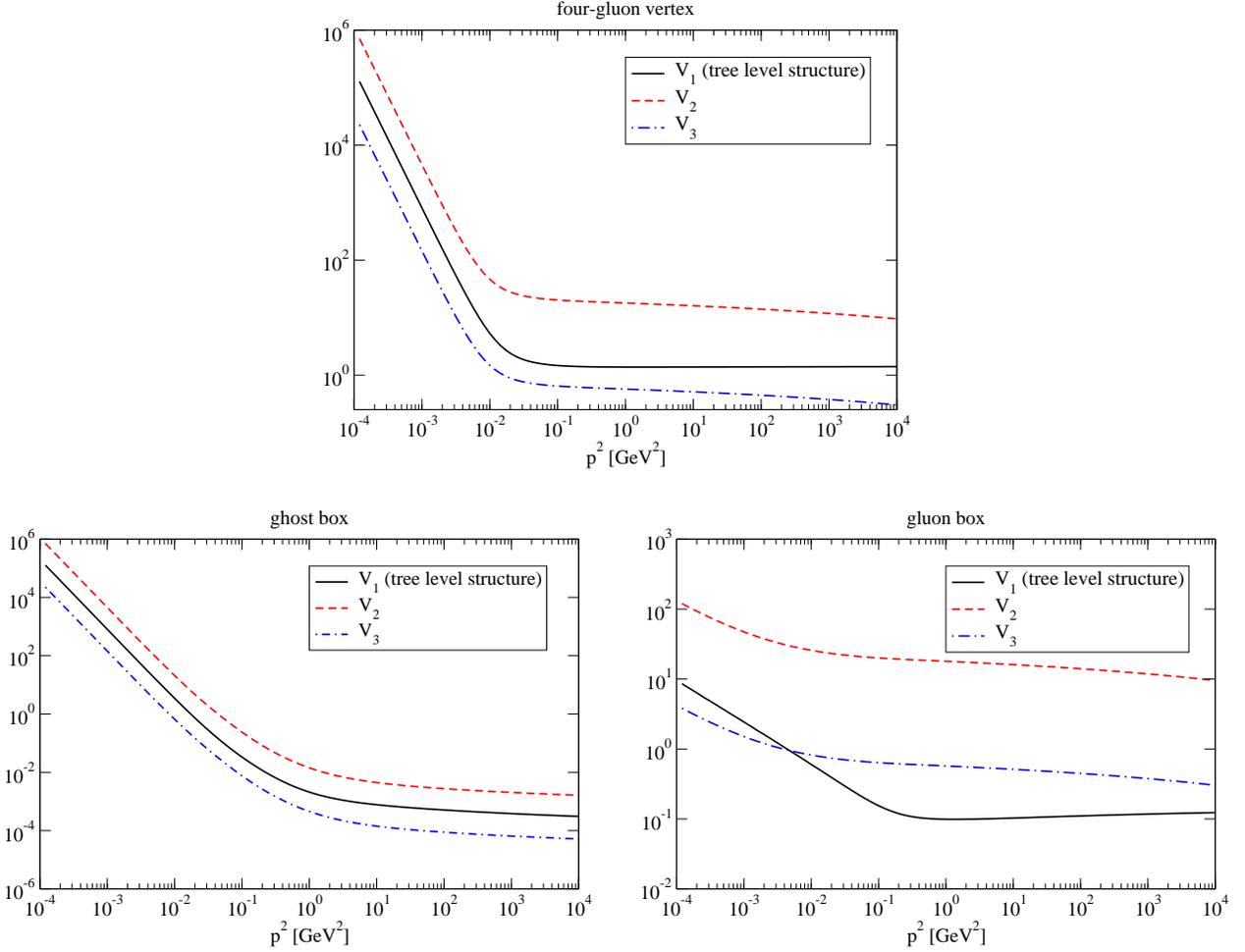

\centerline{\epsfig{file=res_vertex.eps,width=8cm}} \vspace*{5mm}
\centerline{\epsfig{file=res_ghostbox.eps,width=8cm} \hspace*{4mm}
            \epsfig{file=res_gluonbox.eps,width=8cm}}
\caption{Results for the full four-gluon vertex (top) and the ghost 
box diagram (bottom left) and the gluon box diagram (bottom right) 
in the kinematical section $p_1=-3p, p_2=p_3=p_4=p$. The results 
for the ghost and gluon boxes include the symmetry factors. The 
three curves in each diagram correspond to projections onto the 
three Bose-symmetric structures given in eq.~{\ref{4g-structure}}. 
\label{res_ghostbox}}    
\end{figure}

We first present numerical results for the specific kinematical
situation given by 
\begin{center}
\begin{picture}(50,80){
 \put(1,1){\includegraphics[width=3cm,keepaspectratio]{4GluonVertexDressed.eps}}}
 \put(-10,60){$p$}
 \put(-10,35){$p$}
 \put(-10,5){$p$}
 \put(100,35){$-3p$}
\end{picture}
\end{center}
which matches the one used in our infrared and ultraviolet analysis. The
results are presented in Fig.~\ref{res_ghostbox}. On the top diagram 
we display the full four-gluon vertex in this kinematical setup
projected onto the Bose-symmetric tensor structures given in 
eq.~(\ref{4g-structure}). Recall that the structure $V_1$ is identical
to the one of the bare four-gluon vertex ('tree-level structure').
The results match nicely our expectations from the analytical analysis
in section \ref{IR}. All structures of the four-gluon vertex diverge
like $(p^2)^{-4\kappa}$ in the infrared. This divergence is driven
from the ghost-loop diagram, as can be seen from comparing the full
result with the contributions from the ghost box and gluon box diagrams
displayed in the lower panel of Fig.~\ref{res_ghostbox}. Concerning
the infrared coefficients it is not the Bose-symmetric tree-level structure
that dominates but one of the non-tree-level counterparts.

All curves show a characteristic scale of a few hundred MeV, where the
infrared power law behavior bends towards the logarithmic, perturbative
behavior in the ultraviolet momentum region. Certainly, the magnitude of
this scale is inherited from the input (\ref{input}) for the ghost 
and gluon dressing functions and represents the scale $\Lambda_{YM}$ of 
Yang-Mills theory generated by dimensional transmutation.

In the ultraviolet momentum regime we also reproduce the analytic behavior
of the tree-level structure determined in section \ref{IR}. Here the
leading contribution stems from the gluon box diagram. Similar to the
infrared, we also observe that the non-tree-level structure $V_2$
has the largest coefficient of the three structures considered. However,
this will change for even larger momenta, since the logarithms appearing
in $V_2$ and $V_3$ have negative anomalous dimensions, while the tree-level
structure $V_1$ has the correct and positive anomalous dimension $1+4\delta$
in agreement with resummed perturbation theory. 

At first sight, it seems counter-intuitive that the structure $V_2$ dominates
the vertex also for the relatively large momenta considered in our calculations.
However, this dominance has a natural interpretation: it is the prefactors
of these contributions stemming from the corresponding color contractions
that give large relative coefficients of the order of $10^2$ between the
$V_1$ and $V_2$ projections. These also appear in first order perturbation
theory, i.e. with no internal dressings from propagators and vertices in the
ghost and gluon box diagrams. We explicitly checked that the relative ordering
of the contributions $V_1, V_2$ and $V_3$ is the same in this case. This
shows that the ordering appearing in Fig.~\ref{res_ghostbox} is not an artefact 
of the truncation of the ghost-gluon and three-gluon vertices.
The relative magnitudes of $V_1,V_2$ and $V_3$ may however be modified by the
inclusion of the missing one-loop diagrams (c),(d) and (e) of Eq.~(\ref{4g-DSE}).

\begin{figure}[t]
\centerline{\epsfig{file=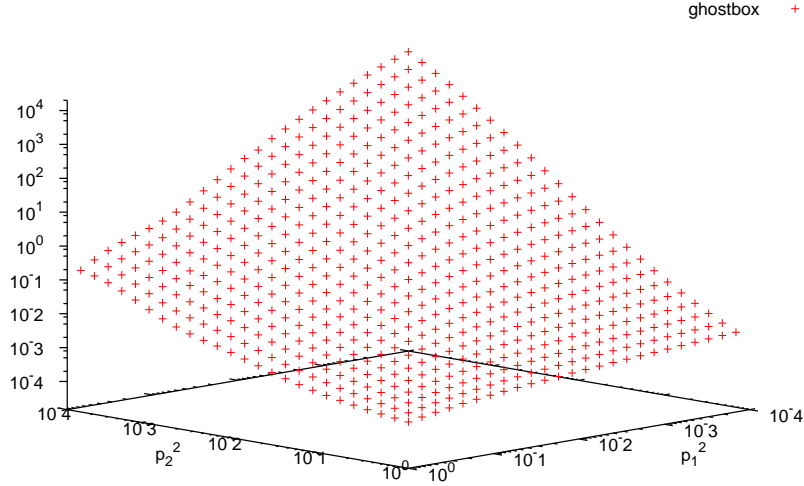,width=12cm}}
\caption{Results for the ghostbox in the kinematical situation with 
$p_1 \cdot p_2=|p_1||p_2|$,
$p_3=p_2$ and $p_4=-p_1-2p_2$ (all four momenta $p_{1..4}$ are defined to flow into 
the diagram).} 
\label{res_multi}    
\end{figure}

In Fig.~\ref{res_multi} we also present a calculation for a different kinematical 
situation with the two independent Lorentz invariants $p_1^2$ and $p_2^2$ and
$p_1 \cdot p_2=|p_1||p_2|$, $p_3=p_2$, $p_4=-p_1-2p_2$ (all four momenta $p_{1..4}$ are defined 
to flow into the diagram), i.e.
\begin{center}
\begin{picture}(50,80){
 \put(1,1){\includegraphics[width=3cm,keepaspectratio]{4GluonVertexDressed.eps}}}
 \put(-10,60){$p_1$}
 \put(-10,35){$p_2$}
 \put(-10,5){$p_2$}
 \put(100,35){$(-p_1-2p_2)$}
\end{picture}
\end{center}
As can be seen from Fig.~\ref{res_multi} we find an infrared divergency when all
momenta go to zero with the power law $(p^2)^{-4\kappa}$ satisfied in the presence
of only one external scale, i.e. in a cone around the diagonal of the diagram. The
behaviour of the vertex dressing function for kinematics at the edges of the
diagram is nontrivial and may indicate additional, weaker kinematical singularities 
present when one or more external momenta are held fixed. Note that the numerical 
problems in these case are quite intricate, since the presence of an additional scale 
involves huge cancellations for some kinematical points. Here we dealt with these 
problems by employing an adaptive framework for the angular integration routines.
Since the main focus of this study is the running coupling defined along the
diagonal of Fig.~\ref{res_multi} we postpone further discussion of the general 
kinematical behaviour of the vertex to future work. 
 
\section{The running coupling \label{sec_coupling}}

\begin{figure}[t]
\centerline{\epsfig{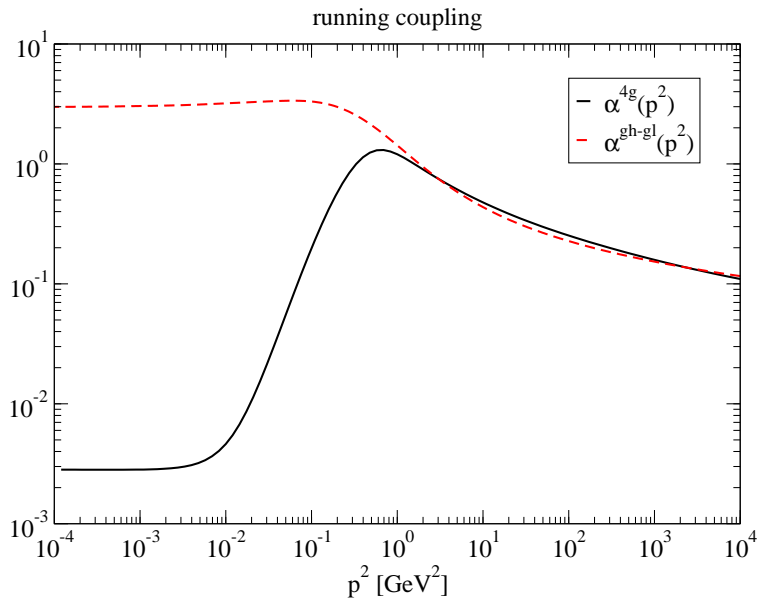}} \vspace*{5mm}
\caption{The running coupling from the four-gluon vertex compared
to the coupling from the ghost-gluon vertex from 
ref.~\cite{Fischer:2002hna}.\label{res_coupling}}    
\end{figure}

The running coupling from the four-gluon vertex has already been given in 
eq.(\ref{alpha-4g}) and is repeated here for the convenience of the reader:
\beqa
\alpha^{4g}(p^2) &=& \frac{g^2}{4 \pi} \, [\Gamma^{4g}(p^2)] \, Z^2(p^2) 
     \,. 
\eeqa
Here $\Gamma^{4g}(p^2)$ denotes the dressing of the tree-level vertex
structure $V_1$ and $Z(p^2)$ denotes the dressing function of the gluon 
propagator. The value $\frac{g^2}{4 \pi}=\alpha(\mu^2)=0.97$ has also been 
given before. The resulting momentum dependence of the coupling is shown
in Fig.~\ref{res_coupling} along with the running coupling from the
ghost-gluon vertex. In the ultraviolet momentum region we observe that
both couplings run like the usual inverse logarithm as well known from
perturbation theory. Here we have universal behavior as dictated from
gauge invariance. In the mid-momentum region we observe a steep rise of
both couplings up to values of order $\alpha \sim 1$. The coupling
from the ghost-gluon vertex then keeps rising before freezing to an
infrared fixed point, whereas the coupling from the four-gluon vertex
decreases dramatically until it reaches a very small but nonzero fixed 
point in the deep infrared. 

Before we discuss the implications of this behavior further, we wish to
verify this numerical result from our analytical calculations in
section \ref{IR}. There we found that 
\begin{eqnarray}
 ^{IR}\Gamma_{gh}\left(p^2\right)
 &\approx& 1.045  \cdot 10^{-4}  \cdot \alpha(\mu^2) \cdot N_c \cdot 
                B^4 \cdot (p^2)^{-4\kappa} \nonumber\\
&=&C_{gh} \cdot \alpha(\mu^2) \cdot N_c \cdot B^4 \cdot \left(p^2\right)^{-4\kappa}.
\end{eqnarray}
In the infrared the gluon dressing function obeys the power-law
\begin{equation}
 Z\left(p^2\right)=A\,\left(p^2\right)^{2\kappa}.
\end{equation}
We then obtain
\begin{eqnarray}
 \alpha_{4g}\left(p^2\rightarrow0\right)&=&
        \alpha\left(\mu^2\right) \cdot C_{gh} 
	\cdot \alpha(\mu^2) \cdot N_c \cdot B^4
	\cdot\left(p^2\right)^{-4\kappa} \cdot
	\left[A\left(p^2\right)^{2\kappa}\right]^2 \nonumber \\
&=&\left[\alpha\left(\mu^2\right) \cdot A \cdot B^2\right]^2 \cdot N_c \cdot C_{gh},
\label{alpha_an}
\end{eqnarray}
This expression is manifestly RG-invariant, since we know from the
coupling of the ghost-gluon vertex that $\alpha(\mu^2) A B^2$ is an
RG-invariant. Also, since $\alpha(\mu^2)=g^2/(4\pi) \sim 1/N_c$ and
$A$ and $B$ are separately independent of 
$N_c$ \cite{von Smekal:1997is,Fischer:2002hna} as is $C_{gh}$, the coupling is 
proportional to $1/N_c$ in agreement with the large $N_c$ counting rules.

Note that the combination $\alpha(\mu^2)\, A\, B^2$ is equivalent 
to the running coupling from the ghost-gluon vertex at zero momentum
\cite{Lerche:2002ep}. Thus we can rewrite eq.~(\ref{alpha_an}) as
\begin{eqnarray}
 \alpha_{4g}\left(p^2\rightarrow0\right)&=& 
 [\alpha_{gh-gl}\left(p^2\rightarrow0\right)]^2 \cdot C_{gh} \cdot N_c
\end{eqnarray}
With $\alpha^{gh-gl}(0) \approx 8.92/N_c$ \cite{Lerche:2002ep} and
$C_{gh} = 1.045  \cdot 10^{-4}$ we then obtain 
\begin{eqnarray}
 \alpha_{4g}\left(p^2\rightarrow0\right) \approx \frac{0.0083}{N_c}. 
\end{eqnarray}
This value agrees well with our numerical result.

In order to assess the implications of our findings it is worth noting
that the corresponding coupling from the three-gluon vertex has a very 
similar behavior than the one of the four-gluon vertex shown in figure
\ref{res_coupling}. In particular it also has an extremely small fixed
point in the infrared, a maximum at intermediate momenta and a perturbative 
logarithmic tail in the ultraviolet \cite{Schwenzer}. We therefore find
universality in the ultraviolet momentum region, as required from gauge 
invariance. In the infrared, all three couplings are qualitatively similar 
in the sense that they all go to an infrared fixed point (as already 
emphasized in \cite{Alkofer:2004it}). However, there are huge qualitative 
differences between the coupling involving ghosts, $\alpha_{gh-gl}$, and 
the other two couplings, $\alpha_{3g}$ and $\alpha_{4g}$, that only 
involve gluonic correlators. 

In this respect it is important to note, that the smallness of the infrared 
fixed point of the four-gluon vertex is rooted in the structure of the 
vertex-DSE. In section (\ref{IR_ghost}) we found, that the diagram involving 
ghosts is the one that gives the leading infrared behavior and determines 
the coefficient of the infrared power law $C_{gh} (p^2)^{-4 \kappa}$ of the 
four-gluon vertex. One reason why the coefficient $C_{gh}$ is small is a 
factor $1/216$ stemming from the projection on the tree-level tensor 
structure. The four propagators in the loop generate further suppression. 
Therefore the smallness of this coefficient can be attributed entirely to 
the structure of the DSE and does not depend on our choice for the ghost-gluon 
and three-gluon vertices.

Of course, the ghost box is only the first term in the skeleton expansion of 
the original ghost related diagram in the full DSE, eq.~(\ref{4g-DSE}). It is 
known from ref.~\cite{Alkofer:2004it} that all terms in this expansion share 
the same infrared power law behavior and will therefore contribute to the 
coefficient $C_{gh}$. Certainly, we cannot exclude that a summation of these 
terms will result in large changes compared to our value of $C_{gh}$.   

\begin{figure}[t]
\centerline{\epsfig{file=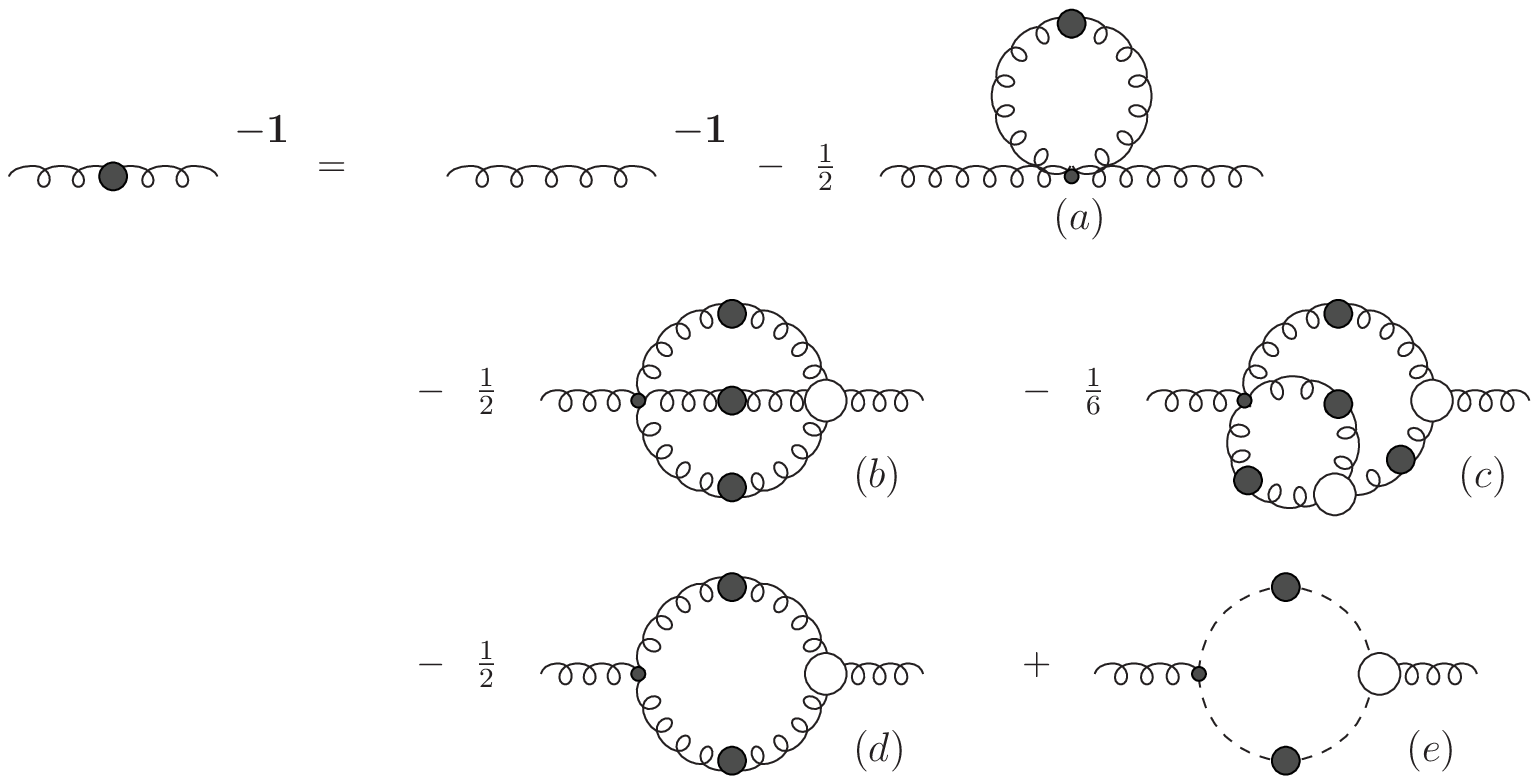,width=12cm}}\vspace*{10mm}
\centerline{\epsfig{file=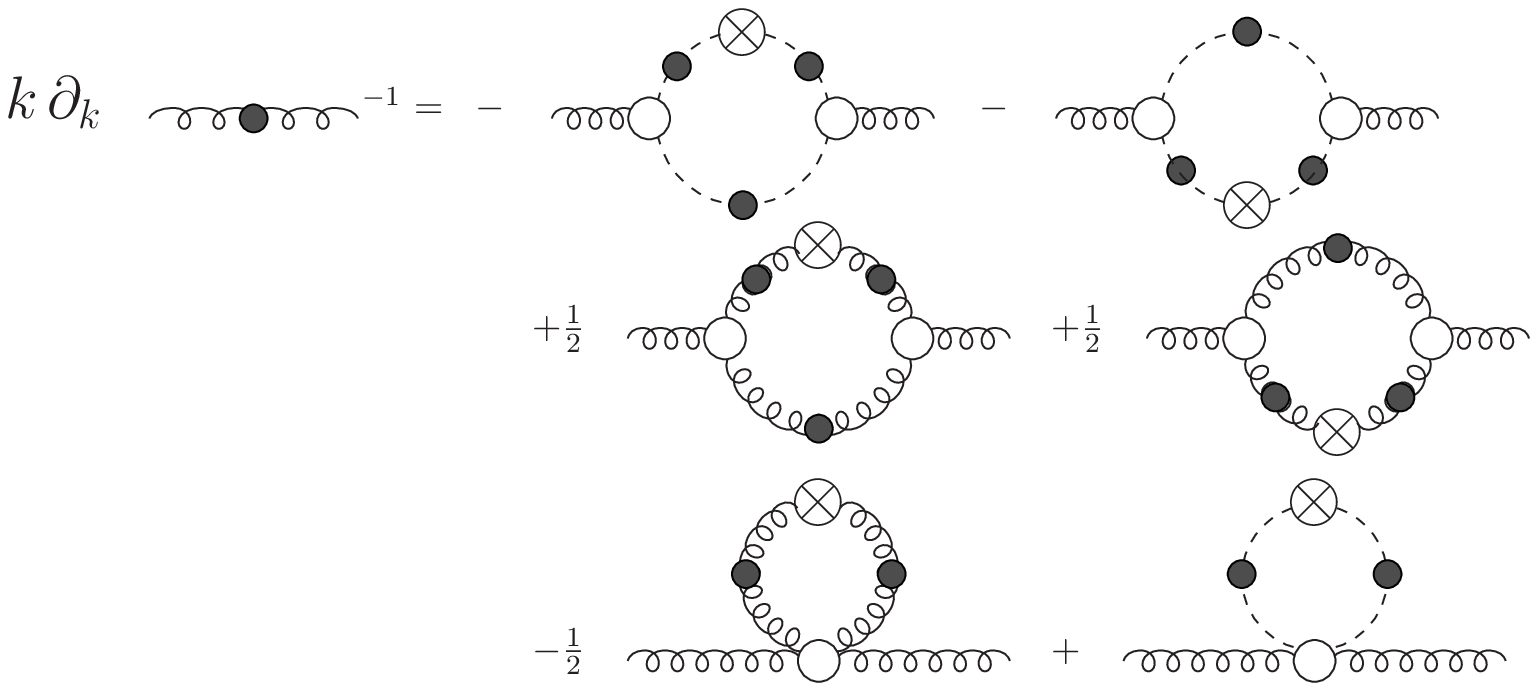,width=12cm}}
\caption{The Dyson-Schwinger equation (top) and the functional flow
equation (bottom) for the gluon propagator. In the flow equation, crosses
denote insertions of the infrared regulator, which cuts off the theory
at or around a scale $k$.\label{DE}}
\end{figure}

However, there is a systematic argument that indicates this may indeed be 
not the case. Consider once again the Dyson-Schwinger equation for the gluon 
propagator. In Fig.~\ref{DE} we compare this equation with the 
corresponding flow equation in the functional renormalization group 
framework. From a systematic point of view, the DSE can be viewed as an 
integrated flow equation, so the physical content of the two equations 
is the same. However, there is an important structural difference between 
the two equations: In every diagram of the DSE we have one bare vertex, 
whereas in the flow equation, all vertices are dressed. As has been 
noted in \cite{Fischer:2006vf} this leads to an interesting situation for 
small momenta: In the DSE the ghost-loop $(a)$ is the only diagram 
responsible for the leading contributions in the infrared, all other 
diagrams are suppressed by powers of momenta. In the flow equation, 
however, all diagrams share the same infrared exponent and therefore 
contribute to the coefficient of the resulting power law for the gluon 
dressing functions. Now, both equations are exact as they stand, so they 
should give the same results in the infrared. We also know that the 
ghost-gluon vertex is almost bare in the small momentum region. For 
this reasons we expect that the ghost-loop diagrams in the flow equations 
(first line) should be roughly similar in effect to the ghost-loop (a) in 
the gluon-DSE \cite{Pawlowski:2003hq,Fischer:2004uk}. Thus the infrared 
coefficients of all other diagrams in the flow equations have to either 
cancel, or should be much smaller than the coefficient of the ghost 
diagrams. Our result for the coupling from the four-gluon vertex together 
with the tentative result for the three-gluon vertex \cite{Schwenzer}  
indicates exactly this: since all these diagrams are roughly proportional 
to their corresponding coupling the gluonic diagrams in the flow equation 
are parametrically suppressed compared to the ghost-diagrams due to the
smallness of the gluonic couplings in the infrared. 
This offers a natural explanation how the DSE and the renormalization 
group framework can agree in the infrared. In turn, this point suggests 
that the smallness of the four-gluon coupling may indeed be an effect
which is robust beyond the leading order in the skeleton expansion.

In fact there is a further argument supporting this scenario. In 
\cite{Zwanziger:2002ia} Zwanziger gave good arguments for an infrared 
effective theory dominated by the Faddeev-Popov determinant. He argued that
all purely gluonic interactions switch off in the infrared and it is the 
geometry of the gauge group which then controls the infrared dynamics via
the ghost content of the theory. This is exactly what we found here. 

\section{Summary \label{summary}}

In this work we investigated the nonperturbative structure of the
four-gluon vertex from (a truncated version of) its Dyson-Schwinger
equation. We identified analytically the leading infrared and
ultraviolet terms of this equation and found good agreement of this
analysis with our numerical solution. We investigated the behavior
of the three Bose symmetric tensor structures that can be constructed
from a subset of the complete tensor basis of the vertex. The dressing
functions of these three structures all show an infrared singular 
behavior with power laws in agreement with the results from naive 
power counting \cite{Alkofer:2004it}. In the ultraviolet momentum 
region our solutions reproduce resummed perturbation theory.

The central result of our work concerns the running coupling
from the four-gluon vertex, built from a combination of vertex dressing
and the dressing function of the gluon propagator. Although in the
ultraviolet momentum region the coupling agrees nicely with the one
from the ghost-gluon vertex (as it should, according to gauge invariance),
in the infrared we observe strong deviations. Whereas the coupling from
the ghost-gluon vertex develops an infrared fixed point at around 
$\alpha_{gh-gl}(0) \approx 9/N_c$, we find a much smaller fixed point
at around $\alpha_{3g}(0) \approx 9 \cdot 10^{-3}/N_c$ for
the coupling from the four-gluon vertex. 

Certainly, the stability of this finding has to be checked wrt further 
improvements of our truncation scheme. These have to include a study of 
the two-loop diagrams in the skeleton expansion of the ghost-part of the 
vertex DSE, since (only) these terms have the potential to change the infrared 
coefficients of the vertex and therefore the value of the infrared fixed 
point. However, on general grounds we are confident that the smallness of 
the running coupling 
from the four-gluon vertex is an important property which is stable wrt 
these improvements. As discussed in the last section, the reason is that
this fact explains why the Dyson-Schwinger and the functional 
renormalization group equations for the two point functions of Yang-Mills
theory agree in the infrared, although their structure is quite different.
With small couplings from the three- and four-gluon vertices gluonic
contributions to the infrared behaviour of the ghost and gluon FRGs are
parametrically suppressed, leading to ghost dominance in agreement with the
results from the DSEs. This finding also supports the notion of an infrared 
effective theory dominated from the Faddeev-Popov determinant proposed in 
\cite{Zwanziger:2002ia}. 

\smallskip
{\bf Acknowledgments}\\
We are grateful to Jan Pawlowski for discussions on the 
comparison between Dyson-Schwinger equations and the functional 
renormalization group equations. We also thank Reinhard Alkofer, 
Markus Huber and Kai Schwenzer for discussions and communicating
preliminary results of ref.~\cite{Schwenzer} to us. This work has been 
supported by the Helmholtz-University Young Investigator Grant VH-NG-332.

\begin{appendix}

 \section{The tensor basis \label{app_B}}
 
The tensor basis is constructed on the following building blocks of 
Lorentz- and color-tensors.\\
Color-tensors:\\
\begin{center}
\begin{tabular}{llll}
$C^{(1)}_{abcd}=\delta_{ab}\delta_{cd},$ & $C^{(2)}_{abcd}=\delta_{ac}\delta_{bd},$ 
& $C^{(3)}_{abcd}=\delta_{ad}\delta_{bc},$ & $C^{(4)}_{abcd}=f_{abn}f_{cdn},$ \\
$C^{(5)}_{abcd}=f_{acn}f_{dbn}$ &&&
\end{tabular}
\end{center}
\vspace{0.5cm}
Lorentz-tensors:\\
\begin{center}
\begin{tabular}{lll}
$L^{\kappa\lambda\mu\nu}_{(1)}=\delta^{\kappa\lambda}\delta^{\mu\nu}$ &
$L^{\kappa\lambda\mu\nu}_{(2)}=\delta^{\kappa\mu}\delta^{\lambda\nu}$ &
$L^{\kappa\lambda\mu\nu}_{(3)}=\delta^{\kappa\nu}\delta^{\lambda\mu}$
\end{tabular}\\
\end{center}
\vspace{0.5cm}
From these, a preliminary generator system of the tensor space can be 
constructed\\
\begin{center}
\begin{tabular}{llll}
$B_{1}=L^{(1)}C_{(1)}$ & $B_{2}=L^{(1)}C_{(2)}$ 
& $B_{3}=L^{(1)}C_{(3)}$ & $B_{4}=L^{(1)}C_{(4)}$ \\
$B_{5}=L^{(1)}C_{(5)}$ & $B_{6}=L^{(2)}C_{(1)}$ 
& $B_{7}=L^{(2)}C_{(2)}$ & $B_{8}=L^{(2)}C_{(3)}$ \\
$B_{9}=L^{(2)}C_{(4)}$ & $B_{10}=L^{(2)}C_{(5)}$ 
& $B_{11}=L^{(3)}C_{(1)}$ & $B_{12}=L^{(3)}C_{(2)}$ \\
$B_{13}=L^{(3)}C_{(3)}$ & $B_{14}=L^{(3)}C_{(4)}$ 
& $B_{15}=L^{(3)}C_{(5)},$ &
\end{tabular}
\end{center}
\vspace{0.5cm}
with the Lorentz/color-indices left implicit. The tree-level tensor-structure 
of the four-gluon vertex
\begin{eqnarray}
V^{(0)} & \propto & f_{abn}f_{cdn} 
(\delta^{\kappa\mu}\delta^{\lambda\nu}-\delta^{\kappa\nu}\delta^{\lambda\mu})
\nonumber \\
&& +f_{acn}f_{bdn}
(\delta^{\kappa\lambda}\delta^{\mu\nu}-\delta^{\kappa\nu}\delta^{\lambda\mu})
\nonumber \\
&& +f_{adn}f_{bcn}
(\delta^{\kappa\lambda}\delta^{\mu\nu}-\delta^{\kappa\mu}\delta^{\lambda\nu}),
\nonumber
\end{eqnarray}
is not a member of the preliminary system. To construct a system containing 
the tree-level structure a Gram-Schmidt-algorithm is applied. It yields an 
orthogonal system of basis-tensors $^{(j)}T_{abcd}^{\kappa\lambda\mu\nu}$.
For projection purposes it is also useful to define normalised quantities
${(j)}U_{abcd}^{\kappa\lambda\mu\nu}$ such that
\begin{eqnarray}
 \left(c\cdot\,^{(j)}T_{abbcd}^{\kappa\lambda\mu\nu}\right)
 \cdot\,^{(k)}T_{abcd}^{\kappa\lambda\mu\nu}&\equiv&
 ^{(j)}U_{abcd}^{\kappa\lambda\mu\nu}
 \cdot\,^{(k)}T_{abcd}^{\kappa\lambda\mu\nu}\nonumber\\
 &=&\delta_{jk}.
\end{eqnarray}
The basis system constructed this way is given by:\\

\begin{tabular}{l}

$U_{(1)}=\frac{1}{108N_c^{2}(N_c^{2}-1)}\,(-B_{4}+2B_{5}+2B_{9}-B_{10}-B_{14}-B_{15})$ \\ \\

$U_{(2)}=\frac{1}{468N_c^{2}(N_c^{2}-1)}\,(-B_{4}+2B_{5}+2B_{9}-B_{10}+5B_{14}-B_{15})$ \\ \\

$U_{(3)}=\frac{1}{3510N_c^{2}(N_c^{2}-1)+\frac{123201}{4}(N_c^{2}-1)^{2}-21060N_c(N_c^{2}-1)}$ \\ \\
$\qquad \qquad \cdot(B_{4}-2B_{5}-2B_{9}+B_{10}+\frac{351}{8}B_{13}+\frac{29}{2}B_{14}+B_{15})$ \\ \\
\end{tabular}

\begin{tabular}{l}

$U_{(4)}=\frac{1}{\frac{276246}{25}N_c^{2}(N_c^{2}-1)+\frac{6472953}{100}(N_c^{2}-1)^{2}-\frac{1657476}{25}N_c(N_c^{2}-1)-\frac{848232}{25}(N_c^{2}-1)}$ \\ \\
$\qquad \qquad \cdot(B_{4}-2B_{5}-2B_{9}+B_{10}+\frac{306}{5}B_{12}-\frac{693}{40}B_{13}-\frac{263}{10}B_{14}+B_{15})$ \\ \\

$U_{(5)}=\frac{1}{5616N_c^{2}(N_c^{2}-1)+\frac{10138203}{25}(N_c^{2}-1)^{2}-33696N_c(N_c^{2}-1)-\frac{4363794}{25}(N_c^{2}-1)}$ \\ \\
$\qquad \qquad \cdot(-8B_{4}+16B_{5}+16B_{9}-8B_{10}+\frac{3141}{20}B_{11}-\frac{369}{20}B_{12}-\frac{369}{20}B_{13}+B_{14}-8B_{15})$ \\ \\

$U_{(6)}=\frac{1}{13816N_c^{2}(N_c^{2}-1)+294(N_c^{2}-1)^{2}-1372N_c(N_c^{2}-1)-294(N_c^{2}-1)}$ \\ \\
$\qquad \qquad \cdot(B_{4}-2B_{5}-2B_{9}+\frac{92}{3}B_{10}-\frac{7}{2}B_{11}+\frac{7}{4}B_{12}+\frac{7}{4}B_{13}-\frac{23}{6}B_{14}+B_{15})$ \\ \\
\end{tabular}

\begin{tabular}{l}

$U_{(7)}=\frac{1}{\frac{345536}{15}N_c^{2}(N_c^{2}-1)+96(N_c^{2}-1)^{2}-448N_c(N_c^{2}-1)-96(N_c^{2}-1)}$ \\ \\
$\qquad \qquad \cdot(\frac{136}{9}B_{4}-\frac{272}{9}B_{5}+\frac{2776}{90}B_{9}-\frac{104}{9}B_{10}-2B_{11}+B_{12}+B_{13}-\frac{1244}{90}B_{14}+\frac{136}{9}B_{15})$ \\ \\

$U_{(8)}=\frac{1}{\frac{1591288}{25}N_c^{2}(N_c^{2}-1)+\frac{10743516}{25}(N_c^{2}-1)^{2}-\frac{9486288}{25}N_c(N_c^{2}-1)-\frac{7776}{25}(N_c^{2}-1)}$ \\ \\
$\qquad \qquad \cdot(-B_{4}+2B_{5}+\frac{846}{5}B_{8}+\frac{193}{5}B_{9}+\frac{178}{5}B_{10}-\frac{18}{5}B_{11}+\frac{9}{5}B_{12}-\frac{81}{2}B_{13}-\frac{68}{5}B_{14}+B_{15})$

\end{tabular}

\begin{tabular}{l}

$U_{(9)}=\frac{1}{\frac{18614232}{125}N_c^{2}(N_c^{2}-1)+\frac{101940444}{125}(N_c^{2}-1)^{2}-\frac{111685392}{125}N_c(N_c^{2}-1)-\frac{48261744}{125}(N_c^{2}-1)}$ \\ \\
$\qquad \qquad \cdot(-B_{4}+2B_{5}+\frac{11304}{50}B_{7}-\frac{2844}{50}B_{8}-\frac{5606}{50}B_{9}+\frac{178}{5}B_{10}$ \\
$\qquad \qquad \quad-\frac{18}{5}B_{11}-\frac{2736}{50}B_{12}+\frac{801}{50}B_{13}+\frac{1204}{50}B_{14}-B_{15})$ \\ \\

$U_{(10)}=\frac{1}{\frac{48940416}{625}N_c^{2}(N_c^{2}-1)+\frac{1828870727}{5625}(N_c^{2}-1)^{2}-\frac{46543936}{125}N_c(N_c^{2}-1)-\frac{48261744}{125}(N_c^{2}-1)}$ \\ \\
$\qquad \qquad \cdot(2B_{4}-4B_{5}+\frac{7074}{50}B_{6}-\frac{1386}{50}B_{7}-\frac{1386}{50}B_{8}+\frac{178}{50}B_{9}-\frac{356}{5}B_{10}$ \\
$\qquad \qquad \quad -\frac{2817}{100}B_{11}+\frac{4}{3}B_{12}+\frac{4}{3}B_{13}-B_{14}+2B_{15})$ \\ \\

$U_{(11)}=\frac{1}{\frac{49098}{5}N_c^{2}(N_c^{2}-1)+1440(N_c^{2}-1)^{2}-6720N_c(N_c^{2}-1)-1440(N_c^{2}-1)}$ \\ \\
$\qquad \qquad \cdot(\frac{107}{6}B_{4}+\frac{25}{6}B_{5}+2B_{6}-B_{7}-B_{8}-\frac{37}{30}B_{9}-\frac{11}{2}B_{10}$ \\
$\qquad \qquad \quad -8B_{11}+4B_{12}+4B_{13}-\frac{129}{10}B_{14}+\frac{107}{6}B_{15})$ \\ \\

$U_{(12)}=\frac{1}{5670N_c^{2}(N_c^{2}-1)+1440(N_c^{2}-1)^{2}-6720N_c(N_c^{2}-1)-1440(N_c^{2}-1)}$ \\ \\
$\qquad \qquad \cdot(\frac{25}{2}B_{4}-\frac{25}{6}B_{5}-2B_{6}+B_{7}+B_{8}-\frac{29}{6}B_{9}+\frac{11}{2}B_{10}$ \\
$\qquad \qquad \quad +8B_{11}-4B_{12}-4B_{13}-\frac{41}{6}B_{14}-\frac{107}{6}B_{15})$ \\ \\

$U_{(13)}=\frac{1}{1080N_c^{2}(N_c^{2}-1)+7290(N_c^{2}-1)^{2}-6480N_c(N_c^{2}-1)}$ \\ \\
$\qquad \qquad \cdot(\frac{45}{2}B_{3}+5B_{4}+5B_{5}-\frac{9}{2}B_{8}-B_{9}-B_{10}-\frac{9}{2}B_{13}-B_{14}-B_{15})$ \\ \\

$U_{(14)}=\frac{1}{2520N_c^{2}(N_c^{2}-1)+13666(N_c^{2}-1)^{2}-14976N_c(N_c^{2}-1)-6048(N_c^{2}-1)}$ \\ \\
$\qquad \qquad \cdot(30B_{2}-\frac{14}{2}B_{3}-15B_{4}+5B_{5}-6B_{7}+\frac{3}{2}B_{8}+3B_{9}-B_{10}$ \\
$\qquad \qquad \quad -6B_{12}+\frac{3}{2}B_{13}+3B_{14}-B_{15})$ \\ \\

$U_{(15)}=\frac{1}{1920N_c^{2}(N_c^{2}-1)+9720(N_c^{2}-1)^{2}-11520N_c(N_c^{2}-1)-6480(N_c^{2}-1)}$ \\ \\
$\qquad \qquad \cdot(25B_{1}-5B_{2}-5B_{3}+\frac{20}{3}B_{4}-\frac{40}{3}B_{5}-5B_{6}+B_{7}+B_{8}-\frac{4}{3}B_{9}+\frac{8}{3}B_{10}$ \\
$\qquad \qquad \quad -5B_{11}+B_{12}+B_{13}-\frac{4}{3}B_{14}+\frac{8}{3}B_{15})$

\end{tabular}

\section{The construction of the Bose symmetric tensor basis \label{app_C}}

To construct a basis of Bose-symmetric tensor-structures out of the structures given 
in appendix \ref{app_B}, one first has to construct a matrix representation of the 
permutation group with respect to the tensor structures. A general tensor in the linear 
space ${\cal V}$ given by the tensor-basis from appendix \ref{app_B} is represented by 
a vector
\begin{eqnarray}
 T&=&\sum_{i=1}^{15}a_i\cdot\,^{(i)}U_{\kappa\lambda\mu\nu}^{abcd}\nonumber \\
&\equiv&(a_1,a_2,\dots,a_{15})
\end{eqnarray}
An important feature of this tensor-basis, is that the tensor-structures 
constructed from it are closed under permutations of the external momenta. 
I.e. no tensor-structures not included from the very beginning are created 
by such permutations. This also means that permutations of the external momenta 
map the tensor space onto itself. Let $\cal P$ be a permutation of the external 
momenta, then
\begin{displaymath}
 v\in{\cal V}\,:\quad {\cal P}v\in{\cal V}.
\end{displaymath}
This means that there is a matrix-representation of the permutation. For the 
four-gluon vertex, there are $4!$ permutations. Let $M_{(j)}$ be the 
matrix-representation for the $j$-th permutation. What one is interested 
in are the vectors $v\in{\cal V}$ that are invariant under {\em all} permutations. 
To find them one first has to find the eigenvectors with respect to the 
eigenvalue one for each member of the matrix-representation of the permutations. 
The eigenvalue one in general can have different geometrical multiplicities for 
each of these matrices. This restricts the dimension of the Bose symmetric tensor 
space. The maximum number of linear independent vectors in the Bose symmetric space 
is the difference of the lowest and the second lowest geometrical multiplicity of 
the eigenvalue one in the set of matrices of the representation of the permutations.

For the four-gluon vertex tensor-structures the lowest geometrical multiplicity 
of the eigenvalue one is five, while the second lowest is eight. Thus one ends up 
with an upper limit of three dimensions for the Bose symmetric tensor space. Having 
calculated the eigenvectors of the representation matrices, one can construct the 
full Bose symmetric linear tensor space. Let $e_{k,l}$ be the $l$-th eigenvector 
with respect to eigenvalue one of the $k$-permutation matrix.
\begin{displaymath}
 M_{(j)}e_{k,l}=e_{k,l}.
\end{displaymath}
Consider two permutations ${\cal P}_1$ and ${\cal P}_2$, with matrix representations 
$M_1$ and $M_2$. Denote the eigenvectors with respect to the eigenvalue one of these 
permutations as $v_i$ and $w_i$ respectively. Let the geometrical multiplicities of 
the eigenvalue one be $\mu_1$ and $\mu_2$. The vectors, that are simultaneously 
included in the eigenspaces of two permutations, are found as the solution of the 
equation
\begin{equation}
 \sum_{i=1}^{\mu_1}\xi_i\cdot v_i=\sum_{j=1}^{\mu_2}\bar \xi_j\cdot w_j.
\label{E2}
\end{equation}
If this coupled system of algebraic equations is determined it can easily be solved 
by standard methods. This is the case, when there is only one eigenvector that is 
invariant under all permutations. The case of an under-determined system also is not 
problematic. The overdetermined system (which occurs in the case of the four-gluon 
tensors) is more complicated. It is then useful to reformulate the problem. Let $A$ 
be the matrix, that is constructed from the column eigenvectors of both permutations 
in the following way
\begin{equation}
 A=\Bigg(\Bigg(v_1^T\Bigg)\,\Bigg(v_2^T\Bigg)\,\dots \,\Bigg(v_{\mu_1}^T\Bigg)\,
	\Bigg(-w_1^T\Bigg)\,\Bigg(-w_2^T\Bigg)\,\dots\,\Bigg(-w_{\mu_2}^T\Bigg)\Bigg).
\end{equation}
With this definition
\begin{equation}
 A\cdot(\xi_1,\dots,\xi_{\mu_1},\bar \xi_1,\dots,\bar \xi_{\mu_2})^T=0
\end{equation}
is equivalent to eq.~(\ref{E2}). The non-trivial kernel of the matrix $A$ consists 
of $\mu_1+\mu_2$-dimensional vectors, whose first $\mu_1$ components are the solution 
for the $\xi_1$ and the others are the solutions for the $\bar \xi_j$. The kernel of 
the matrix can then be evaluated using standard methods. For the four-gluon vertex 
the result is given in the main body of this work, eq.~(\ref{4g-structure}).

\section{Decompositions of higher Green's functions\label{app_D}}

In the following the decompositions of the reducible vertices in eq.~(\ref{4g-DSE}) 
into one-particle irreducible vertices will be given \cite{Driesen:1998xc,Driesen:1997wz}.
The reducible functions are denoted as $T$ and the irreducible as $\Gamma$.

\begin{enumerate}
 \item The decomposition of the four gluon reducible Green's function:
  \begin{equation}
   \parbox{3cm}{
   \begin{picture}(30,80){
	  \put(0,0){\includegraphics[width=3cm,keepaspectratio]{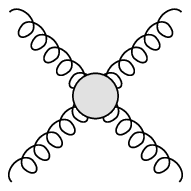}}
	  \put(38,38){\colorbox{mygray}{\scriptsize $T$}}
	  }
	 \end{picture}
	 }=
   \parbox{3cm}{
   \begin{picture}(30,80){
	  \put(0,0){\includegraphics[width=3cm,keepaspectratio]{4GlueAmp1.eps}}
	  \put(38,38){\colorbox{mygray}{\scriptsize $\Gamma$}}
	  }
	 \end{picture}
	 }+
   \parbox{3cm}{
   \begin{picture}(30,80){
	  \put(0,0){\includegraphics[width=3cm,keepaspectratio]{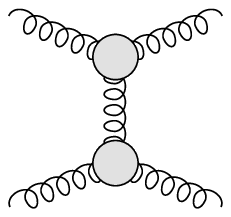}}
	  \put(38,57){\colorbox{mygray}{\scriptsize $\Gamma$}}
	  \put(38,16){\colorbox{mygray}{\scriptsize $\Gamma$}}
	  }
	 \end{picture}	  
	 }+
   \parbox{3cm}{
   \begin{picture}(30,80){
	  \put(0,0){\includegraphics[width=3cm,keepaspectratio]{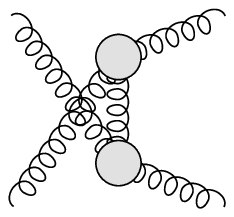}}
	  \put(39,57){\colorbox{mygray}{\scriptsize $\Gamma$}}
	  \put(39,16){\colorbox{mygray}{\scriptsize $\Gamma$}}
	  }
	 \end{picture}
	 }
  \end{equation}
 \item The decomposition of the five gluon reducible function:
  \begin{eqnarray}
   \parbox{2.5cm}{
   \begin{picture}(25,80){ 
	  \put(0,0){\includegraphics[width=2.5cm,keepaspectratio]{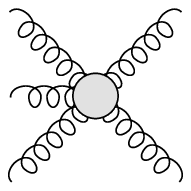}}
	  \put(30,30){\colorbox{mygray}{\scriptsize $T$}}
	  }
	 \end{picture}
	 }&=&
   \parbox{2.5cm}{
   \begin{picture}(25,80){
	  \put(0,0){\includegraphics[width=2.5cm,keepaspectratio]{5GlueAmp1.eps}}
	  \put(30,30){\colorbox{mygray}{\scriptsize $\Gamma$}}
	  }
	  \end{picture}
	  }+
   \parbox{3cm}{
   \begin{picture}(30,80){
	  \put(0,0){\includegraphics[width=3cm,keepaspectratio]{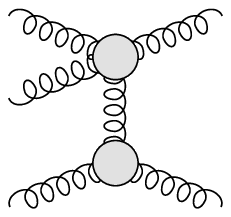}}
	  \put(38,57){\colorbox{mygray}{\scriptsize $\Gamma$}}
	  \put(38,16){\colorbox{mygray}{\scriptsize $\Gamma$}}
	  }
	 \end{picture}
	 }+
   \parbox{3cm}{
   \begin{picture}(30,80){
	  \put(0,0){\includegraphics[width=3cm,keepaspectratio]{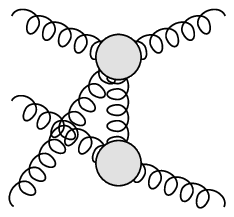}}
	  \put(38,57){\colorbox{mygray}{\scriptsize $\Gamma$}}
	  \put(38,16){\colorbox{mygray}{\scriptsize $\Gamma$}}
	  }
	 \end{picture}	  
	 }\nonumber \\
&&\qquad\qquad+
   \parbox{3cm}{
   \begin{picture}(30,80){
          \put(0,0){\includegraphics[width=3cm,keepaspectratio]{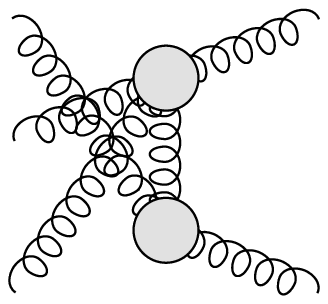}}
          \put(38,57){\colorbox{mygray}{\scriptsize $\Gamma$}}
	  \put(38,16){\colorbox{mygray}{\scriptsize $\Gamma$}}
	  }
	 \end{picture}
	 }+
   \parbox{3cm}{
   \begin{picture}(30,80){
	  \put(0,0){\includegraphics[width=3cm,keepaspectratio]{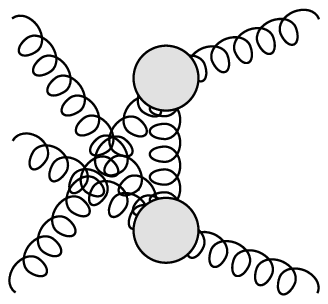}}
          \put(38,57){\colorbox{mygray}{\scriptsize $\Gamma$}}
	  \put(38,16){\colorbox{mygray}{\scriptsize $T$}}
	  }
	 \end{picture}	  
	 }\nonumber \\
&&\qquad\qquad+
   \parbox{3cm}{
   \begin{picture}(30,80){
	  \put(0,0){\includegraphics[width=3cm,keepaspectratio]{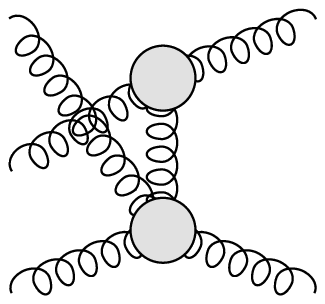}}
	  \put(38,57){\colorbox{mygray}{\scriptsize $\Gamma$}}
	  \put(38,16){\colorbox{mygray}{\scriptsize $T$}}
	  }
	 \end{picture}
	 }+
   \parbox{3cm}{
   \begin{picture}(30,80){
	  \put(0,0){\includegraphics[width=3cm,keepaspectratio]{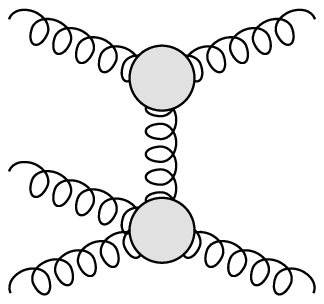}}
	  \put(38,57){\colorbox{mygray}{\scriptsize $\Gamma$}}
	  \put(38,16){\colorbox{mygray}{\scriptsize $T$}}
	  }
	 \end{picture}
	 }
  \end{eqnarray}

\end{enumerate}

\section{Analytical integration methods}

\subsection{Collected integrals}
 \begin{enumerate}
  \item From ref.\cite{Gradshteyn1965} we use:
	\begin{enumerate}
	 \item \begin{eqnarray*}
	        \int_0^{\pi}\frac{\sin^{2\mu-1}x\,dx}{\left(1+2a\cos x+a^2\right)^{\nu}}&=&
			B\left(\mu,\frac{1}{2}\right)F\left(\nu,\nu-\mu+\frac{1}{2};\mu+\frac{1}{2};a^2\right) 
\nonumber \\ \hspace{1cm} \nonumber\\ \left[\text{Re}\,\mu >0,|a|<1\right]\hspace{1.5cm}&&\hspace{10cm}
	       \label{C1}
	       \end{eqnarray*}
	       \begin{equation}
 		\hspace{1cm}
	       \end{equation}
	\item \begin{eqnarray*}
	       \int_0^1(1-x)^{\mu-1}x^{\nu-1}\,_pF_q(a_1,\dots,a_p;b_1,\dots,b_q;ax)dx= \hspace{5cm} \\
		=\frac{\Gamma(\mu)\Gamma(\nu)}{\Gamma(\mu+\nu)}\,_{p+1}F_{q+1}(\nu,a_1,\dots,a_p;\mu+\nu,
			b_1,\dots,b_q;a)\hspace{2cm}  \\
		\hspace{1cm}\\
		\left[\text{Re}\,\mu>0,\text{Re}\,\nu>0,p\leq q+1, \text{if}\,p=q+1, \text{then}\,|a|<1\right]
			\hspace{5cm} 
	       \label{C2}
	      \end{eqnarray*}
	      \begin{equation}
 		\hspace{1cm}
	      \end{equation}
	\end{enumerate}
  \item By partial integration, it can be seen:
	\begin{eqnarray*}
	 \int_0^{\pi}d\theta \cos\theta\times\frac{\sin^{2\mu-1}\theta}{\left(1+2a \cos \theta+a^2\right)^{\nu}}&=&
	  -\frac{a\nu}{\mu}\int_0^{\pi}d\theta\frac{\sin^{2\mu+1}}{\left(1+2a\cos\theta+a^2\right)^{\nu+1}} \\
&=&	  -\frac{a\nu}{\mu}B\left(\mu+1,\frac{1}{2}\right)F\left(\nu+1,\nu-\mu+\frac{1}{2};\mu+
		\frac{3}{2};a^2\right)\\
	  \hspace{1cm} \\
		\left[\text{Re}\,\mu >-1,|a|<1\right]\hspace{2cm}
	       \label{C3}
	\end{eqnarray*}
	\begin{equation}
 	 \hspace{1cm}
	\end{equation}	
 \end{enumerate}
 
 \subsection{Integrating with the Chebyshev-expansion \label{app-cheby}}
 A continuous function can be expanded in a series of polynomials. Most commonly one expands the function
 in a Taylor-series. But the convergence properties of a Taylor-series can fail to be sufficiently good, since the error of the approximation
 can be concentrated in a special region of the considered integration interval. When dealing with integrals
 on finite intervals, the Chebyshev-expansion can be an alternative. It has the advantage that the approximation error is smeared out over the   
 interval. The integral of the original function is reduced to integrals over Chebyshev-polynomials.\\ \\
 The Chebyshev-polynomials are
 \begin{equation}
  T_n(x)=\cos(n\arccos(x)).
 \end{equation}
 A function $f(x)$ can be expanded over these polynomials
 \begin{equation}
 f(x)\approx\sum_{j=1}^{N-1} c_j T_j(x)-\frac{c_0}{2},
 \end{equation}
 with the coefficients
 \begin{equation}
 c_j=\frac{2}{N}\sum_{k=1}^{N}\cos\left(\frac{j(k-1/2)\pi}{N}\right)
	f\left[\cos\left(\frac{(k-1/2)\pi}{N}\right)\right],
 \end{equation}
 where $N$ is the order of the expansion. The abscissas $\cos\left(\frac{(k-1/2)\pi}{N}\right)$ are the
 zeros of the n-th Chebyshev polynomial. For a more detailed discussion see \cite{Bloch:1995dd}.\\
 To integrate a function using its Chebyshev-expansion, one transforms the variable, so that
 the integral is on the interval [-1,1].
 \begin{eqnarray}
  \int_a^bdx\,f(x)&\stackrel{x=\tau(y)}{=}&\frac{d\tau}{dy}\int_{-1}^1dy\sum_{j=0}^{N-1}c_jT_j(y)-\frac{c_0}{2}
	\nonumber \\
&=& \frac{d\tau}{dy}\left(-c_0+\sum_{j=0}^{N-1}c_j\int_{-1}^1dy\,T_j(y)\right).
 \end{eqnarray}
 Thus the integration has been reduced to an integration over Chebyshev-polynomials. The integral over the 0-th
 Chebyshev-polynomial yields $2$, while the integral over the 1st vanishes. \\ The integral over the j-th
 Chebyshev polynomial yields
 \begin{equation}
  \int_{-1}^1dx\,T_j(x)=\frac{\cos(j\pi)+1}{1-j^2}.
 \end{equation}
 Thus one obtains
 \begin{equation}
  \int_a^bdx\,f(x)\approx\frac{d\tau}{dy}\left(c_0+\sum_{j=2}^{N-1}c_j\frac{\cos(j\pi)+1}{1-j^2}\right).
 \end{equation}
 Plugging in eq. (D.6) and denoting $y_k=\cos\left(\frac{(k-1/2)\pi}{N}\right)$ one finally gets
 \begin{eqnarray}
 \int_a^bdx\,f(x)\approx\frac{2}{N}\frac{d\tau}{dy}\left(\sum_{k=1}^Nf(y_k)+\sum_{j=2}^{N-1}\sum_{k=1}^N
	\cos\left(\frac{j(k-1/2)\pi}{N}\right)f(y_k)\frac{\cos(j\pi)+1}{1-j^2}\right).
 \end{eqnarray}

 \subsection{The UV Integral}
 
 In the UV analysis of the Ghost- and the Gluon-Box integral of the form
 \begin{equation}
  I\equiv\int_x^{\infty}dy\frac{\left(\omega\log\left(\frac{y}{s}\right)+1\right)^a}{y^n}
 \end{equation}
 occur. To evaluate them, one substitutes $z=\omega\log\left(\frac{y}{s}\right)+1$ yielding
 \begin{equation}
  I=\frac{e^{\frac{n-1}{\omega}}}{\omega s^{n-1}}\int_{z(x)}^{\infty}dz\,e^{-\frac{(n-1)z}{\omega}}z^a
 \end{equation}
 and then $\xi=\frac{(n-1)z}{\omega}$. One gets
 \begin{eqnarray}
  I&=&\frac{e^{\frac{n-1}{\omega}}\omega^a}{s^{n-1}(n-1)^{a+1}}\int_{\xi(z(x))}^{\infty}d\xi \, e^{-\xi}\xi^a 
	\nonumber \\
 &=&	\frac{e^{\frac{n-1}{\omega}}\omega^a}{s^{n-1}(n-1)^{a+1}}
		\Gamma\left(a+1,(n-1)\left(\log\left(\frac{x}{s}\right)+\frac{1}{\omega}\right)\right).
 \end{eqnarray}
 Since $x$ is large in the region of interest, one can employ the asymptotic expansion
 \begin{equation}
  \Gamma(a,x)=x^{a-1} \, e^{-x} \left[\sum_{m=0}^{M-1}\frac{(1-a)_m}{(-x)^m}+O(|x|^{-M})\right],
 \end{equation}
 with the Pochhammer-symbol $(a)_m$
 \begin{equation}
  (a)_0=1, \quad (a)_n=\frac{\Gamma(a+n)}{\Gamma(a)}=a(a+1)\ldots(a+n-1).
 \end{equation}
 Keeping only the first order of the series, one results in
 \begin{equation}
  \int_x^{\infty}dy \, \frac{\left( \omega \log\ \left(\frac{y}{s}\right)+1 \right)^a}{y^n}\approx
	\frac{\left(\omega \log \left(\frac{x}{s}\right)+1\right)^a}{(n-1) x^{n-1}}.
	\label{C16}
 \end{equation}

\end{appendix}

\end{document}